\documentclass{aa}  

\usepackage{graphicx}
\usepackage{txfonts}
\usepackage{float}
\usepackage{parskip}
\usepackage{adjustbox}
\usepackage[switch]{lineno}
\begin{document}

   \title{A study in scarlet - II. Spectroscopic properties of a sample of Intermediate Luminosity Red Transients}


   \authorrunning{Valerin et al.}
   \author{G. Valerin\inst{1}\thanks{E--mail: giorgio.valerin@inaf.it},
          A. Pastorello\inst{1},
          E. Mason\inst{2},
          A. Reguitti\inst{1,3},
          S. Benetti\inst{1},
          Y.-Z. Cai\inst{4,5,6},             
          T.-W. Chen\inst{7},         
          D. Eappachen\inst{8,9}, 
          N. Elias-Rosa\inst{1,10},
M. Fraser\inst{11}, 
A. Gangopadhyay\inst{12,13},  
E. Y. Hsiao\inst{14},           
D. A. Howell\inst{15,16},
C. Inserra\inst{17},             
L. Izzo\inst{18,19},
J. Jencson\inst{20},         
E. Kankare\inst{21},
R. Kotak\inst{21}, 
P. Lundqvist\inst{12,22},
P. A. Mazzali\inst{23,24},         
K. Misra\inst{25}, 
G. Pignata\inst{26},
S. J. Prentice\inst{27},           
D. J. Sand\inst{28},        
S. J. Smartt\inst{29,30},
M. D. Stritzinger\inst{31}, 
L. Tartaglia\inst{32},
S. Valenti\inst{33},
J. P. Anderson\inst{34,35},
J. E. Andrews\inst{28},
R. C. Amaro\inst{28},              
C. Barbarino\inst{12},
S. Brennan\inst{12},
F. Bufano\inst{36},          
E. Callis\inst{12},
E. Cappellaro\inst{1},
R. Dastidar\inst{37,35},
M. Della Valle\inst{18,38},               
A. Fiore\inst{1,39,40},          
M. D. Fulton\inst{30},
L. Galbany\inst{10,41},
M. Gromadzki\inst{42},
T. Heikkilä\inst{21},            
D. Hiramatsu\inst{14,15,43,44},             
E. Karamehmetoglu\inst{12,31},    
H. Kuncarayakti\inst{21,45},
G. Leloudas\inst{46},  
M. Limongi\inst{47,48,49}
M. Lundquist\inst{28},                
C. McCully\inst{15},
T. E. Müller-Bravo\inst{10,41}, 
M. Nicholl\inst{30},
P. Ochner\inst{1,50}, 
E. Padilla Gonzalez\inst{15,16},    
E. Paraskeva\inst{51},
C. Pellegrino\inst{52},     
A. Rau\inst{53},
D. E. Reichart\inst{54},
T. M. Reynolds\inst{20,55,56},      
R. Roy\inst{57},                                    
I. Salmaso\inst{1},    
M. Shahbandeh\inst{58,59},
M. Singh\inst{60},
J. Sollerman\inst{12},
M. Turatto\inst{1},               
L. Tomasella\inst{1},                                 
S. Wyatt\inst{28},
D. R. Young\inst{30}
          }

   \institute{Affiliations are listed after the acknowledgements
             }

   \date{Received ----; accepted ----}

 
  \abstract
   {}
   {We investigate the spectroscopic characteristics of Intermediate Luminosity Red Transients (ILRTs), a class of elusive objects with peak luminosity between that of classical novae and standard supernovae. Our goal is to provide a stepping stone in the path to unveil the physical origin of such events, thanks to the analysis of the datasets collected.}
   {We present the extensive optical and near-infrared (NIR) spectroscopic monitoring of four ILRTs, namely NGC 300 2008OT-1, AT 2019abn, AT 2019ahd and AT 2019udc. First we focus on the evolution of the most prominent spectral features observed in the low resolution spectra, then we discuss more in detail the high resolution spectrum collected for NGC 300 2008OT-1 with the Very Large Telescope equipped with UVES. Finally we analyse late time spectra of NGC 300 2008OT-1 and AT 2019ahd through comparisons with both synthetic and observed spectra.}
   {Balmer and Ca lines dominate the optical spectra, revealing the presence of slowly moving circumstellar medium (CSM) around the objects. The line luminosity of H$\alpha$, H$\beta$ and Ca II NIR triplet presents a double peaked evolution with time, possibly indicative of interaction between fast ejecta and the slow CSM. The high resolution spectrum of NGC 300 2008OT-1 reveals a complex circumstellar environment, with the transient being surrounded by a slow ($\sim$30 km s$^{-1}$) progenitor wind. At late epochs, optical spectra of NGC 300 2008OT-1 and AT 2019ahd show broad ($\sim$2500 km s$^{-1}$) emission features at $\sim$6170$\AA$ and $\sim$7000$\AA$ which are unprecedented for ILRTs. We find that these lines originate most likely from the blending of several narrow lines, possibly of iron-peak elements.}
   {}


   \maketitle
%
\section{Introduction}

The transient sky is brimming with a wide variety of phenomena, and spectroscopic observations are a crucial tool to navigate through such a complex landscape. Indeed, remarkable efforts have been made in order to promptly classify newly discovered transients, culminating in the formation of dedicated spectroscopic surveys such as the ``Public European Southern Observatory Spectroscopic Survey of Transient Objects" (PESSTO) consortium \citep{SmarttPesto2015}.
The swift collection of spectra is rewarded with key insights on the ongoing physical processes which could not be otherwise probed. In the past years there has been a growing number of studies on objects displaying narrow and intense emission lines coupled with a slow spectral evolution: these are the hallmarks of optically thick circumstellar medium (CSM) surrounding the transient (e.g. \citealt{BlinnikovHandbook}). Such a scenario has been investigated primarily in Supernovae IIn (SNe IIn): terminal stellar explosions within an opaque CSM formed during the years prior to the event \citep{Filippenko1997}. 
Critically, SNe IIn are not the only events showing evidence of optically thick CSM. Violent eruptions of Luminous Blue Variables (LBVs) mimic the characteristics of a genuine SN, despite being non-terminal events: for this reason they earned the designation of ``SN impostors"  (e.g. \citealt{VanDyk_Impostors}). 

The picture was further complicated by the discovery of other classes of transients that display clear signs of CSM in their spectra. Luminous Red Novae (LRNe) have been argued to arise from mergers of non-degenerate stars after a common envelope phase where the final product of the merger event is expected to survive \citep{PastoRedNovae2019}. Finally, perhaps the most poorly understood among the objects discussed so far, there are the subjects of the present study: the Intermediate Luminosity Red Transients (ILRTs), whose nature is still debated. Together, LBVs, LRNe and ILRTs make up a wider class of objects with luminosity intermediate between Classical Novae and standard SNe, called ``Gap Transients" (see \citealt{PastoGapTransients} and \citealt{CaiReviewGap2022} for a review).
While their physical origin may be remarkably different, all these transients have one thing in common: the presence of optically thick CSM surrounding the system.
The resulting similarity among the spectra of the Gap Transients and SNe IIn makes a proper classification of a newly discovered transient quite difficult: it can be challenging to unmask an impostor \citep{Ransome2021Classification}.

Photometric follow-up can provide some important clues for this type of investigation. The light curve shape can be a useful indicator for classification: LBVs alternate periods of quiescence with erratic and powerful outbursts \citep{HumpreysLBV}, LRNe often show distinctive double peaked light curves, while ILRTs are characterized by SN-like light curves \citep{PastoGapTransients}. 
It is still essential to complement the photometric follow-up with a spectroscopic study to characterize each class, despite the partial degeneracy in the properties of these transients.
From a spectroscopic point of view, ILRTs can be summarily described as a black body continuum with narrow emission lines superimposed. Strong Balmer emission lines dominate their spectra together with several Ca features: Ca II H\&K in absorption, strong Ca II NIR triplet in emission as well as the [Ca II] forbidden doublet. Opposed to LRNe, ILRTs show no sign of molecular bands in the late phases of their evolution \citep{PastoRedNovae2019}.


A key difference between ILRTs and SNe IIn is that both the peak observed temperature (T$_{ILRT}$$\sim$7500 K) and peak luminosity (L$_{ILRT}$ $\lesssim$ 5x10$^{41}$ erg s$^{-1}$) of ILRTs are significantly lower compared to SNe IIn, which instead tend to be hot (T$_{IIn}$$\sim$10000 K, \citealt{AverageIInTaddia}, but up to $\sim$15000 K, e.g. \citealt{Reguitti2021foa}) and even brighter than average SNe II (e.g. \citealt{Kiewe2012SneIInBlue}). The photospheric temperature and luminosity observed for SNe IIn can be sustained even for years (e.g. \citealt{Tartaglia2015da}), thanks to the interaction between fast ejecta and slow moving CSM, a mechanism that can efficiently convert kinetic energy into radiation \citep{Smith_Handbook2017}. When the interaction region between ejecta and CSM is visible, specific spectral features can be identified, most notably a blue shoulder on the H$\alpha$ line \citep{TaddiaStrongInteraction}. ILRTs, however, do not show clear signs of strong dynamical interaction between ejecta and CSM in their spectra: their low temperature and luminosity could imply a rather weak ejecta-CSM interaction, or at least that such interaction is hidden underneath layers of unshocked CSM.


As previously mentioned, the physical mechanism behind ILRTs is still debated. Since the photosphere lies within an optically thick gas shell, it is not possible to probe the material underneath the CSM, which could reveal key information on the nature of the transient itself. This uncertainty has led to a variety of interpretations regarding their true nature.
Some researchers have advocated outbursts of post-main sequence stars \citep{Kashi2010Outburst, Hump2011}, while others have suggested that ILRTs are the result of jets originating from interacting binary stars \citep{Soker2021Jets}. Finally, ILRTs were also associated with Electron Capture Supernovae (ECSNe, e.g. \citealt{Nomoto1984,Kitaura2006}) originating from Super AGB stars \citep{Botticella2009,Pumo2009, Cai2021ILRTs}.
ECSNe are expected to have some distinctive features that separates them from classic Fe core collapse events. As was discussed for SN 2018zd, nebular spectra of ECSN are expected to show Ni features stronger than Fe lines, due to the different nucleosynthesis yields in the two scenarios \citep{Hiramatsu2018zd,Callis_E_2018zd}. Indeed, \cite{JerkstrandModels} provide synthetic nebular spectra of a 9 M$_{\odot}$ Fe core collapse SN, and approximate the behaviour of an ECSN by enhancing the Ni content in their Fe core collapse model. The Ni enriched spectra display particularly strong [Ni II] emission lines.
Until now, such an analysis has not been performed for ILRTs due to the lack of high quality nebular spectra, as well as the lack of detected features beside the ones originating in the CSM.

In this paper we present and analyse spectra of four ILRTs, following the photometric analysis of the same targets performed in ``A study in scarlet I. Photometric properties of a sample of Intermediate Luminosity Red Transients" (Valerin et al. 2024, hereafter "Paper I"). The work is structured in the following manner: in Section 2 the data reduction techniques are described. In Section 3  we present the optical and near infrared (NIR) data collected. In Section 4 the line luminosity and velocity are measured and compared within the sample. Section 5 is dedicated to the study of the high resolution spectroscopy obtained for NGC 300 2008OT-1, while in Section 6 we discuss the broad features appearing in the late time spectra of NGC 300 2008OT-1 and AT 2019ahd. Finally, in Section 7 we summarize our work and draw the conclusions.

\section{Data Reduction}
The analysis presented in the following sections is based on the spectroscopic observations of four different ILRTs over the course of several months. Our follow-up campaigns are focused on the optical and Near-Infrared (NIR) domains, employing several telescopes equipped with different instruments. The data reduction procedure is based on the Image Reduction and Analysis Facility (\texttt{IRAF}, \citealt{TodyIRAF1986,TodyIRAF1993}), a software specifically designed for image processing developed within the astronomical community.
Every image undergoes the same preliminary reduction procedure: trimming, bias and flat-field correction. NIR spectra go through an additional sky subtraction step when combining the different nodding positions.
A one dimensional spectrum of the target is extracted from the two dimensional image, while simultaneously performing a background subtraction in order to avoid contamination from the sky emission lines and the host galaxy signal.
The one dimensional spectrum is then calibrated in wavelength on standard spectral lamps, typically showing He, Ne, Ar, Hg and Kr lines.

These procedures are repeated on the spectrum of a standard star (belonging to the list of ESO spectrophotometric standards\footnote{https://www.eso.org/sci/observing/tools/standards/spectra/stanlis.html}), which is routinely obtained during every observation run. Thanks to the comparison between the observed and tabulated flux values of a standard star, it is possible to accurately flux calibrate the spectrum of the target. Standard star observations are also useful to remove the telluric absorption features which affect ground-based observations. An additional, wavelength independent flux calibration is performed, ensuring that the fluxing of the spectrum is compatible with the values measured from coeval photometric observations. Finally, each spectrum is corrected for redshift and reddening, in particular assuming a reddening law R$_{V}$ = 3.1 both for Galactic and internal extinction \citep{CardelliExtinction1989}. The distance, reddening and redshift adopted for each transient are summarized in Table \ref{tab_Dist_Redd}: for a discussion on the choice of the values used, see Paper I.


\begin{table*}
  \centering
  \begin{tabular}{lcccc}
  \hline
    Transient & Host Galaxy & Redshift & Distance [Mpc] &  A$_{V}$ [mag] \\
    \hline
     AT 2019abn & M51 &  1.54 (-) x10$^{-3}$  & 8.6 (0.1) &  2.44 (0.06) \\
     AT 2019ahd & NGC 3423 &  3.35 (0.01) x10$^{-3}$  &  11.1 (0.7) &  0.45 (0.03) \\
     AT 2019udc & NGC 0718 &  5.78 (0.03) x10$^{-3}$  & 19.9 (1.4) & 0.100 (0.001)  \\
     NGC 300 OT & NGC 300 &  4.80 (0.03) x10$^{-4}$ & 1.92 (0.14) & 0.81 (0.06) \\
    \hline
  \end{tabular}
  \caption{Summary of the key information regarding the transients in the sample, including their redshift, distance and total absorption along the line of sight. Errors, when available, are reported in brackets.}
  \label{tab_Dist_Redd}
\end{table*}

\begin{figure*}
\centering
  \makebox[\columnwidth][c]{\includegraphics[width=1.9\columnwidth]{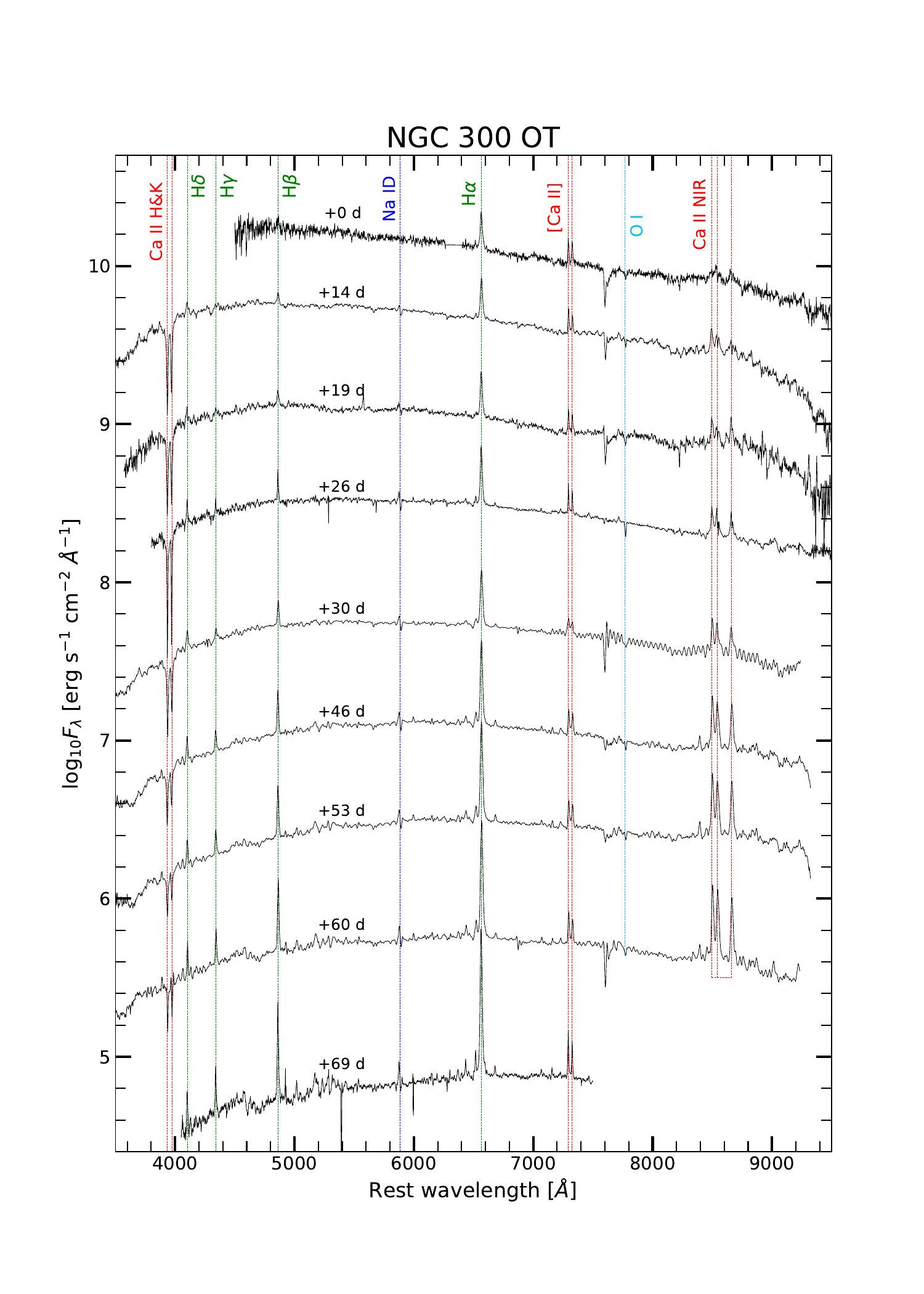}}%
  \caption{NGC 300 OT spectra during the early phases. Ca lines are highlighted in red, H lines in green and, O lines in light blue the Na ID absorption doublet in blue. Epochs are reported with respect to maximum $r$ band luminosity.}
  \label{NGC300OTSpec1}
\end{figure*}

\begin{figure*}
  \centering
  \makebox[\columnwidth][c]{\includegraphics[width=1.9\columnwidth]{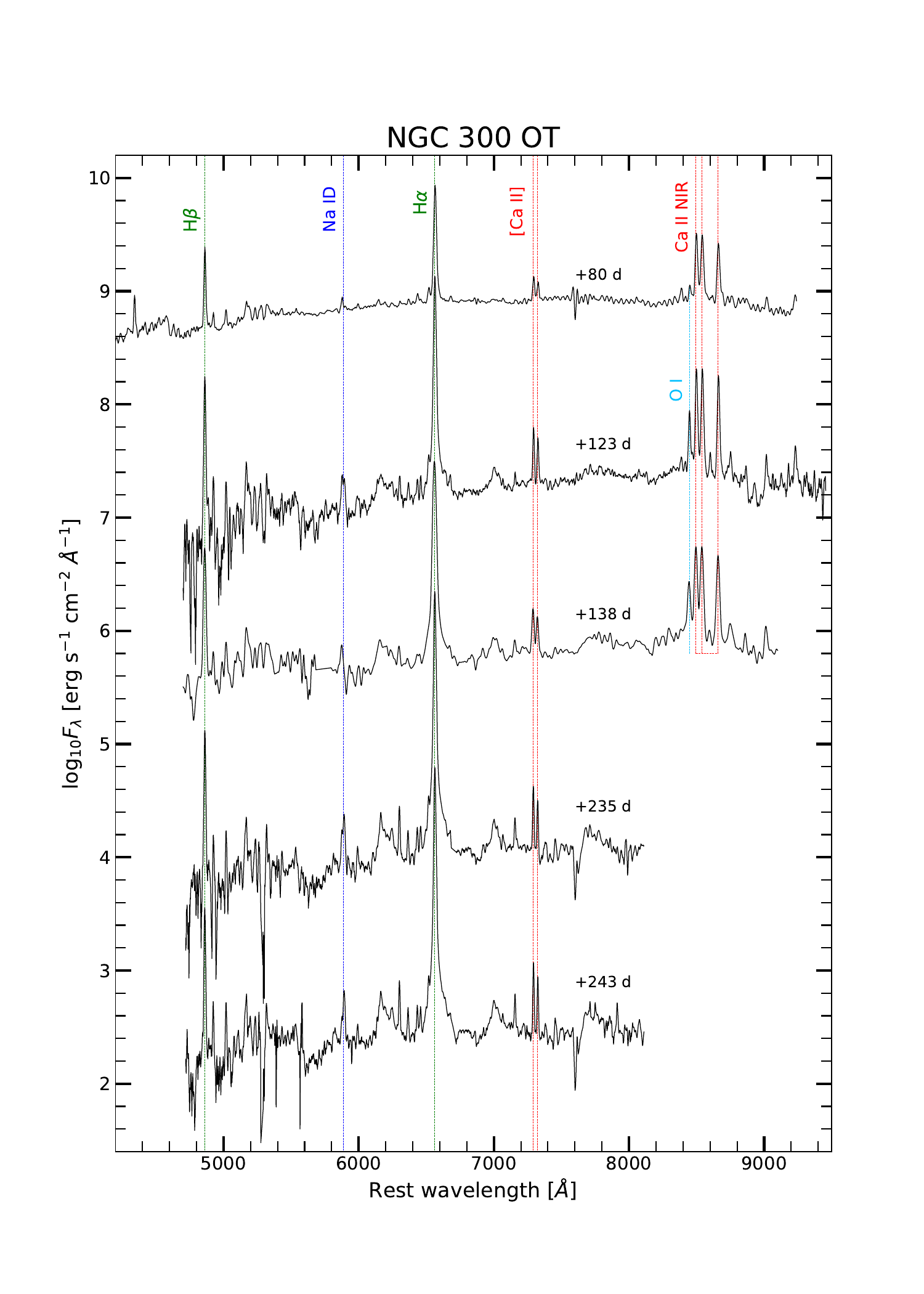}}%
  \caption{NGC 300 OT spectra during the late phases of its evolution. Ca lines are highlighted in red, H lines in green and, O lines in light blue the Na ID absorption doublet in blue. Epochs are reported with respect to maximum $r$ band luminosity.}
  \label{NGC300OTSpec2}
\end{figure*}

Many facilities and instruments were involved in the follow-up campaigns of our targets. Indubitably, the backbone of our spectroscopic monitoring was obtained through the Nordic Optical Telescope (NOT) within the ``NOT Unbiased Transient Survey 2" collaboration (NUTS2, \citealt{Holmbo2019ATEL}) and the instruments available within the ``Global Supernova Project" (GSP, \citealt{AndyLCO}). Other telescopes used during this project include the Gran Telescopio Canaria equipped with OSIRIS, the William Herschel Telescope equipped with ACAM and LIRIS, the South African Large Telescope equipped with RSS, and the New Technology Telescope plus EFOSC2, thanks to the ePESSTO+ consortium \citep{SmarttPesto2015}. Additionally, multiple spectra of NGC 300 OT were taken with the 6.5 m Magellan Telescopes, and the 4.0-m R-C Spectrograph at the Cerro Tololo Inter-American Observatory, and with the Very Large Telescope, most notably a high resolution spectrum with UVES \citep{UVES} and a late spectrum with X-Shooter \citep{X-Shooter}. We also integrated our dataset with some publicly available spectra to have a more complete sampling of each event (\citealt{2019ahdClass} for AT 2019ahd and \citealt{Karambelkar2023} for AT 2019abn).
The detailed logs listing all the spectroscopic observations for each target are reported in Tables  \ref{SpecLog2019abn}, \ref{SpecLog2019ahd}, \ref{SpecLog2019udc} and \ref{SpecLog}.

%

\begin{figure*}
\centering
  \makebox[\columnwidth][c]{\includegraphics[width=1.9\columnwidth]{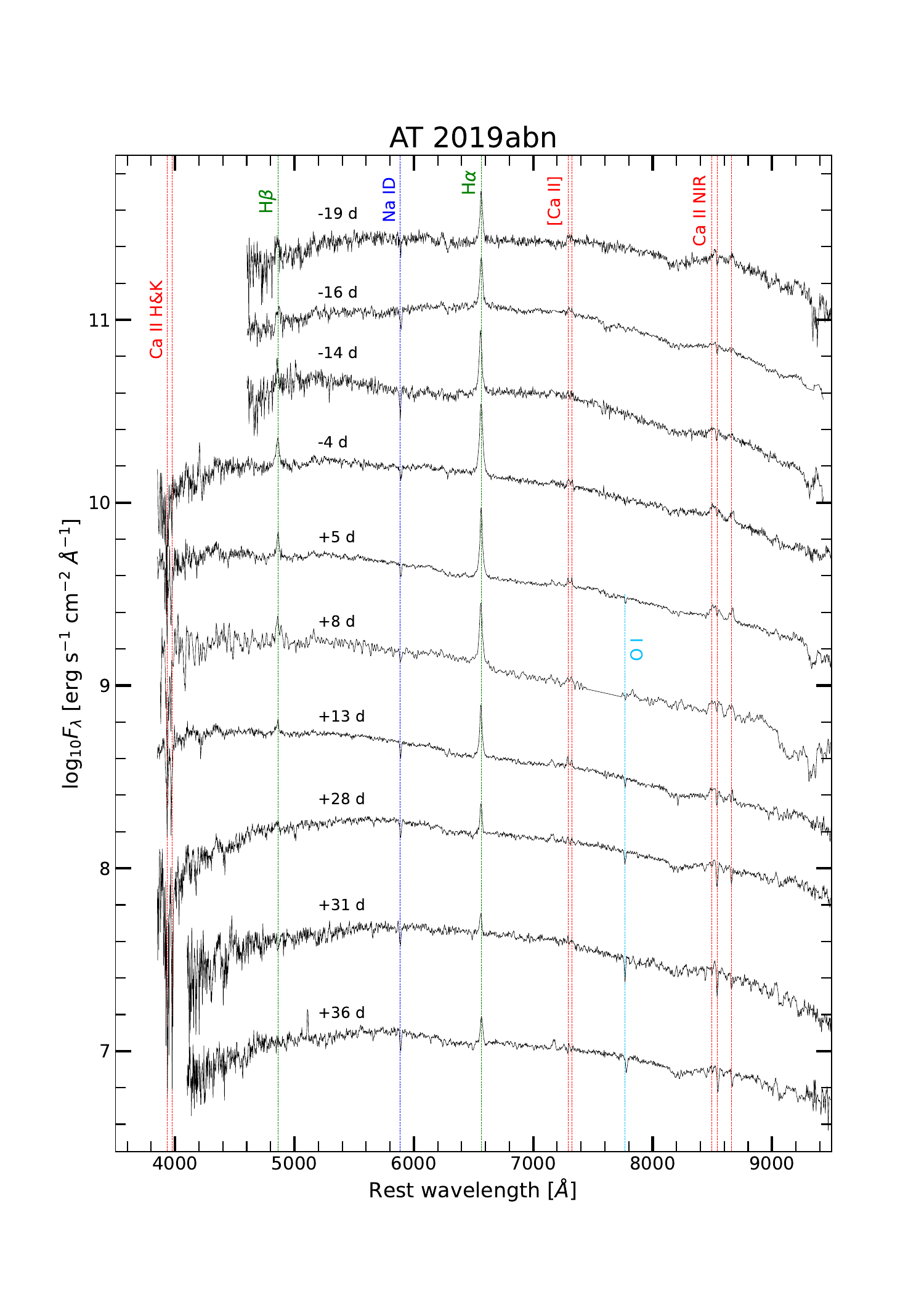}}%
  \caption{AT2019abn spectra collected in the early phases. Ca lines are highlighted in red, H lines in green and, O lines in light blue the Na ID absorption doublet in blue. Epochs are reported with respect to maximum $r$ band luminosity.}
  \label{AT2019abn1}
\end{figure*}


\begin{figure*}
\centering
  \makebox[\columnwidth][c]{\includegraphics[width=1.9\columnwidth]{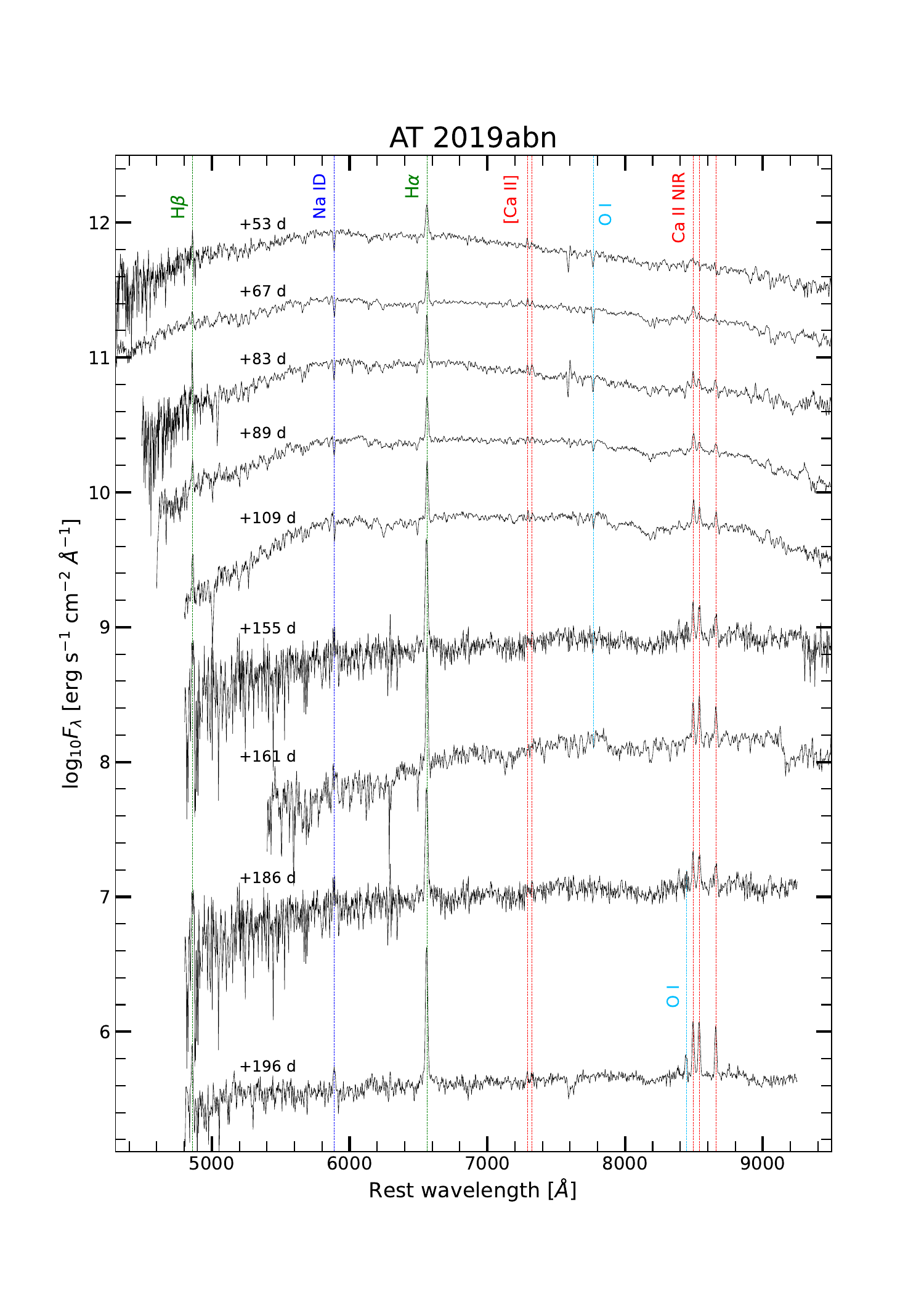}}%
  \caption{AT2019abn late spectra. Ca lines are highlighted in red, H lines in green and, O lines in light blue the Na ID absorption doublet in blue. Epochs are reported with respect to maximum $r$ band luminosity.}
  \label{AT2019abn2}
\end{figure*}


\section{Overview of the data}

\subsection{Optical Spectra} \label{OverviewOpticalSpec}
\begin{figure*}
  \centering
  \makebox[\columnwidth][c]{\includegraphics[width=1.9\columnwidth]{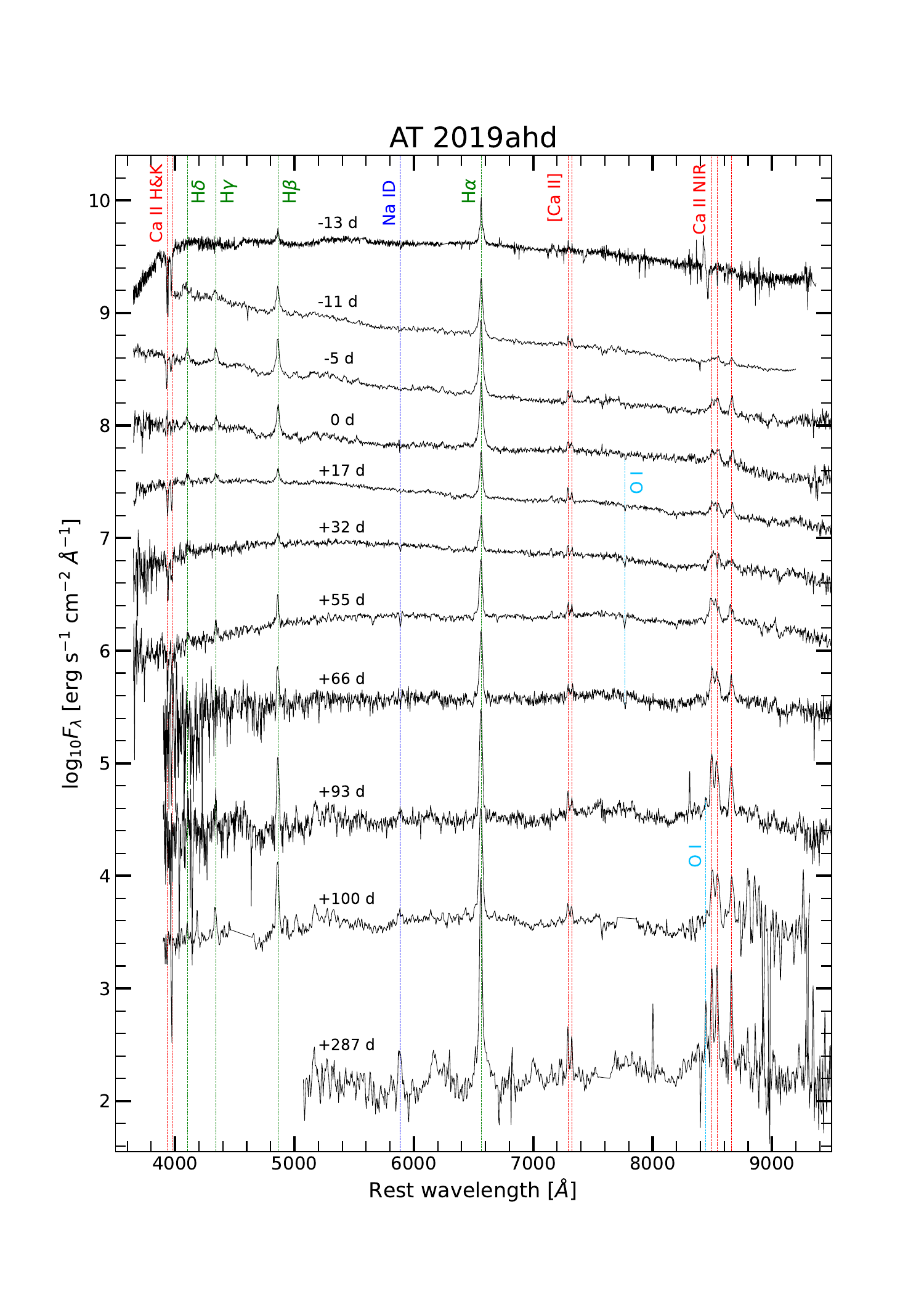}}%
  \caption{Spectra of AT2019ahd. Ca lines are highlighted in red, H lines in green and, O lines in light blue the Na ID absorption doublet in blue. Epochs are reported with respect to maximum $r$ band luminosity.}
  \label{AT2019ahd}
\end{figure*}

\begin{figure*}
  \centering
  \makebox[\columnwidth][c]{\includegraphics[width=1.9\columnwidth]{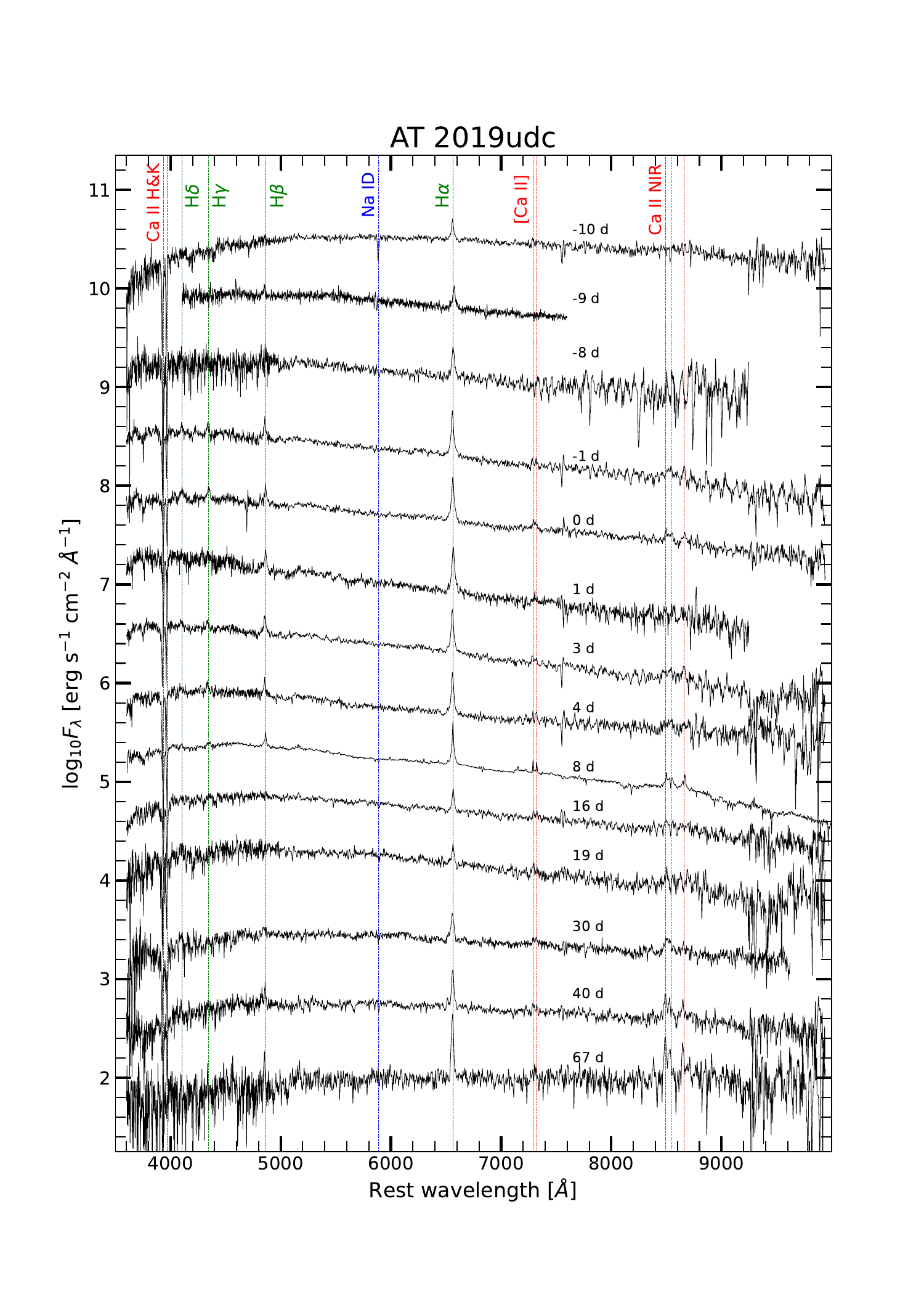}}%
  \caption{Spectra collected for AT2019udc. Ca lines are highlighted in red, H lines in green and, O lines in light blue the Na ID absorption doublet in blue. Epochs are reported with respect to maximum $r$ band luminosity.}
  \label{AT2019udc}
\end{figure*}

 Spectra obtained for NGC 300 OT are displayed in Figures \ref{NGC300OTSpec1} and \ref{NGC300OTSpec2}, while Figures \ref{AT2019abn1} and \ref{AT2019abn2} show the spectral sequence of AT 2019abn. Figures \ref{AT2019ahd} and \ref{AT2019udc} present the spectral evolution of AT 2019ahd and AT 2019udc, respectively. All spectra were corrected for redshift and total extinction along the line of sight: values adopted for these corrections are reported in Table \ref{tab_Dist_Redd}. The most prominent features noticeable in these objects are the emission lines of the Balmer series. Their narrow profiles ($\lesssim$1000 km s$^{-1}$) indicate that the spectral lines are forming in a region where the material is moving slowly, compatible with an opaque CSM. We do not detect clear P-Cygni profiles in the emission lines, which would be caused by absorbing gas approaching the observer along the line of sight. The absence of the P-Cygni features may imply that the photosphere of the transient is located at the outer edge of the CSM, that the transient is intrinsically asymmetric, or that the covering fraction of the absorbing material is low.
The Ca II H \& K absorption lines ($\lambda\lambda$ 3934,3968) are found in all objects, especially at early phases. The photons absorbed through the Ca II H \& K resonance lines excite the electrons of the Ca II atoms from the ground state $3s^{2} 3p^{6} 4s$ $^{2}S_{1/2}$ to $3s^{2} 3p^{6} 4p$ $^{2}P^{o}_{1/2,3/2}$. From this energy level the electrons can radiatively decay to the level $3s^{2} 3p^{6} 3d$ $^{2}D_{3/2,5/2}$, giving rise to a triplet of lines ($\lambda\lambda\lambda$ 8498,8542,8662). Indeed this Ca II NIR triplet is detectable at most epochs in ILRTs, and it becomes dominant at later times, as displayed in Figures \ref{NGC300OTSpec2} and \ref{AT2019abn2}. The compatible shape of the Ca II NIR and Ca II H\& K lines in velocity space corroborates this interpretation (e.g. see Figure \ref{LineHDprofiles}).



\begin{figure*}
 \begin{center}
\makebox[2.0\columnwidth][c]{\includegraphics[width=0.9\textwidth]
  {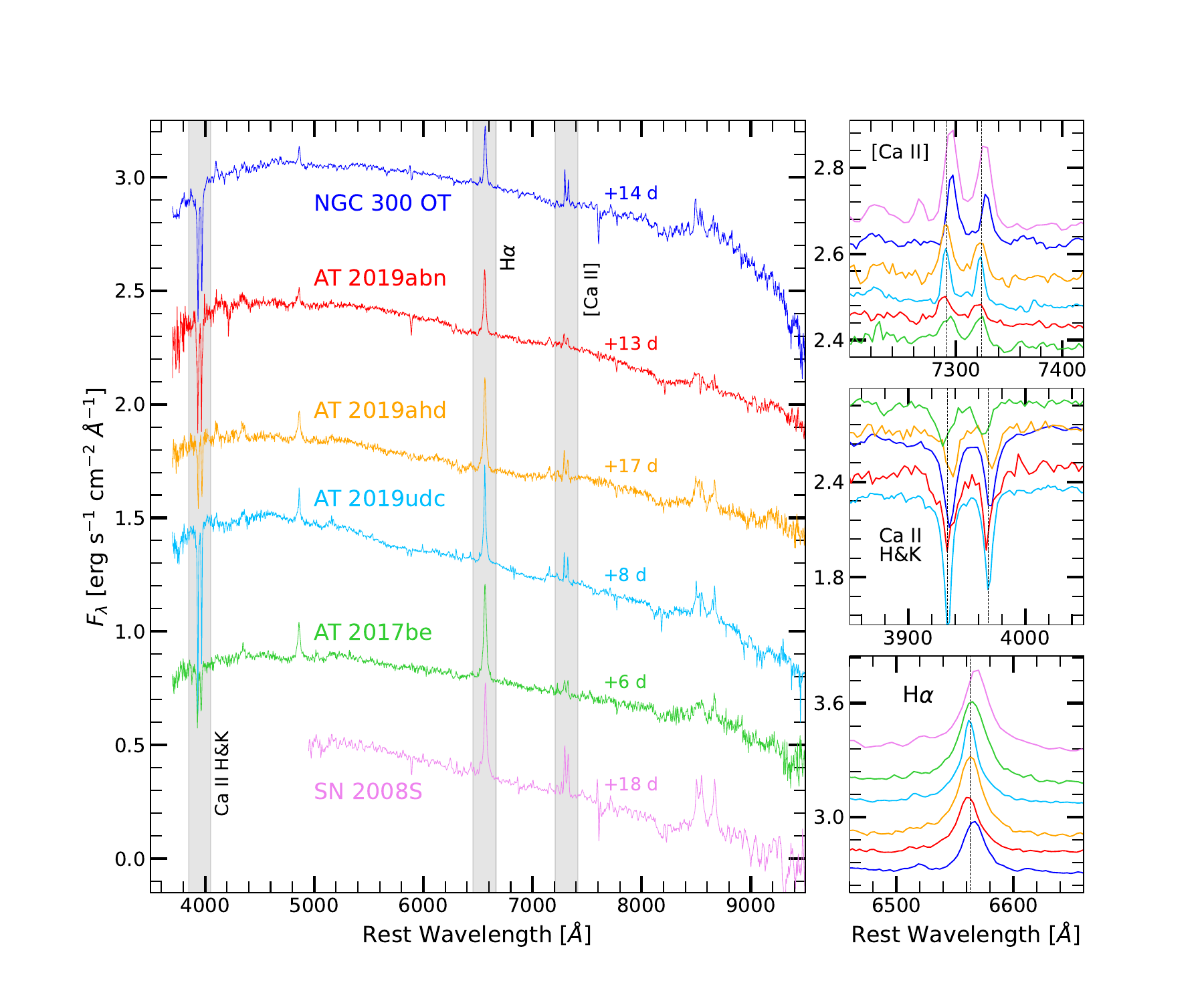}}%
  \caption{Comparison between early spectra of our sample of ILRTs, taken in the first weeks after peak luminosity (phases are reported with respect to maximum light). On the right panels is shown a more detailed view of the [Ca II], Ca II H\&K and H$\alpha$ lines.}
  \label{ComparisonSpectra}
   \end{center}
\end{figure*}

\begin{figure*}
 \begin{center}
  \makebox[2.0\columnwidth][c]{\includegraphics[width=0.9\textwidth]
  {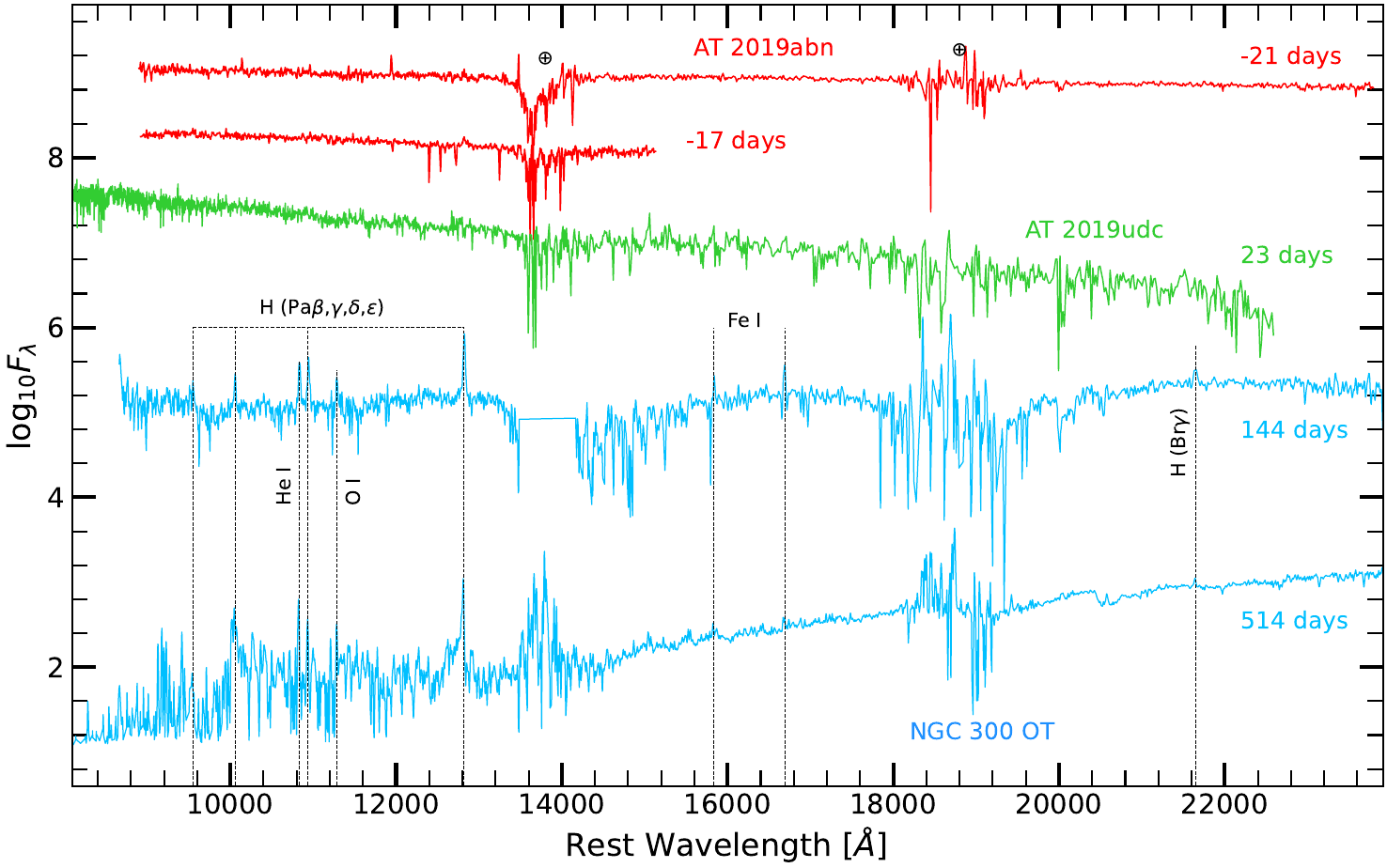}}%
  \caption{Sequence of NIR spectra collected for our sample of ILRTs. In red AT 2019abn, in green AT 2019udc and in blue NGC 300 OT. The most prominent emission lines are marked on the spectra of NGC 300 OT. The position of telluric features is marked with a circled plus symbol. The first spectrum of AT 2019abn displayed is the combination of two spectra obtained at phases -21 and -20 days with respect to maximum.}
  \label{NIRSpectra}
   \end{center}
\end{figure*}

 Finally, the transition that closes this ``loop" envisions the de-excitation of the electrons from the level $3s^{2} 3p^{6} 3d$ $^{2}D_{3/2,5/2}$ back to the ground state $3s^{2} 3p^{6} 4s$ $^{2}S_{1/2}$. Such transition, which produces the [Ca II] doublet ($\lambda\lambda$7291,7323), is forbidden.
 Even if the metastable energy level $3s^{2} 3p^{6} 3d$ $^{2}D_{3/2,5/2}$ is populated, the formation of [Ca II] requires a large amount of gas at moderate excitation and most of all at low density (e.g. \citealt{Merrill_1958}). In fact, in a dense environment, collisions with other atoms or ions de-excitate the electrons back to the ground state before the radiative decay process can take place, consequently suppressing the formation of these lines (e.g. \citealt{Hump2011}). Therefore, this spectral feature is only found in regions where the gas density is below the critical density N$_{crit}$ $\sim$ 10$^{7}$ cm$^{-3}$ (for temperatures typically found in ILRTs, adopting the collision strength reported by \citealt{MelendezCollisionCaII}).
 Based on these premises, the ubiquitous presence of the [Ca II] doublet is one of the most characteristic features in ILRTs spectra. While not unique to ILRTs, the persistent presence of the [Ca II] forbidden doublet in all the objects analysed so far has been proposed as a key feature to distinguish ILRTs from other transients that present similar spectra \citep{PastoGapTransients}. 

 Another noteworthy feature in the low resolution optical spectra of our sample of ILRTs is the presence of O I lines. In the spectra of all the considered objects the O I triplet $\lambda\lambda\lambda$ 7771.9,7774.2,7775.4 is found in absorption. This triplet is detected at multiple epochs in NGC 300 OT (Figure \ref{NGC300OTSpec1}), AT 2019abn (Figures \ref{AT2019abn1} and \ref{AT2019abn2}) and AT 2019ahd (Figure \ref{AT2019ahd}), while in the case of AT 2019udc it is confidently detected only in the GTC spectrum obtained on 2019-11-22, 8 days after maximum. The O I triplet in absorption is a sign that the metastable level 2s$^{2}$2p$^{3}$ ($^{4}$S$^{o}$)3s$^{5}$S$^{o}_{2}$ is populated. We do not detect any emission lines from transitions that may feed this energy level: first of all the $\lambda\lambda\lambda$ 7771.9,7774.2,7775.4 triplet is never clearly detected in emission at any epoch, but we also do not detect transitions from higher energy levels (e.g. O I $\lambda$3947). This suggests that the electrons populated the metastable energy level s$^{2}$2p$^{3}$ ($^{4}$S$^{o}$)3s$^{5}$S$^{o}_{2}$ by recombination after an initial excitation. With time, the O I absorption triplet becomes less pronounced, until its detection becomes uncertain in the latest phases. The absence of a clear emission component of the O I triplet could be due to efficient absorption by electrons in the metastable energy level.
 Few months after maximum luminosity there is also the appearance of the O I $\lambda$8446 emission line, which becomes progressively more prominent with time. In AT 2019abn this feature appears around +155 d, in AT 2019ahd around +93 d and in NGC 300 OT around +123 d (but this emission line is detected already in the high S/N spectrum at +47 d discussed in Section \ref{SECTION_highresspec}). This feature is not found in the spectra of AT 2019udc, but its spectroscopic monitoring had to be stopped at +67 d, therefore this emission line could have developed after the end of our observational campaign. The appearance of the O I $\lambda$8446 line is particularly interesting, since it cannot be the result of the de-excitation of O I atoms, which would result in the coeval formation of the $\lambda\lambda\lambda$ 7771.9,7774.2,7775.4 triplet in emission. Instead, as discussed above, the O I triplet has only been detected in absorption.
 We conclude that the O I $\lambda$8446 emission line found in ILRTs is the consequence of UV pumping. 
 This process is also known as Bowen fluorescence, and it is caused by the resonance between the Ly$\beta$ $\lambda$1025.72 $\AA$ and the O I $\lambda$1025.76 line, which excites the electrons from the ground level 2p$^{4}$ $^{3}$P$_{2}$ to 2p$^{3}$3d $^{3}$D$^{0}_{3}$ \citep{Osterbrock2006}. The electrons subsequently decay to the levels 2p$^{3}$3p $^{3}$P$_{2}$ (emitting $\lambda$11287 $\AA$), 3p$^{3}$3s $^{3}$S$^{0}_{1}$ (emitting $\lambda$8446 
 $\AA$) and finally back to the ground state (through the UV triplet $\lambda\lambda\lambda$1302.2,1304.9,1306.0). Indeed, in the NIR spectra of NGC 300 OT presented in the following section we detect the O I $\lambda$11287 $\AA$ emission line, corroborating the fact that in these ILRTs we are observing the fluorescence process described above, rather than a different pumping mechanism (e.g. see the different processes described by \citealt{Shore2010}). Since the excitation of the O I electrons to the 2p$^{3}$3d $^{3}$D$^{0}_{3}$ energy level requires Ly$\beta$ photons, the appearance of prominent O I $\lambda$8446 $\AA$ at relatively late phases (and not at early ones) needs to be addressed. In the context of faster ejecta expanding underneath a slow moving CSM, the Ly$\beta$ could be shifted off resonance due to Doppler effect: in fact, to have a perfect wavelength match, the Ly$\beta$ photons should be redshifted by 11.7 km s$^{-1}$ with respect to the oxygen atoms. 
 In the early phases, when the resonant lines are opaque, the Ly$\beta$ escapes the ejecta from a compact pseudophotosphere which is moving faster than the CSM above, greatly reducing the coupling between the Ly$\beta$ and the O I $\lambda$1025.76 line. As time progresses and the ejecta expand, the opacity of the lines drops and CSM is finally reached by the redshifted Ly$\beta$ photons that can trigger the Bowen fluorescence, hence explaining the delayed appearance of the O I $\lambda$8446 line. 

 

 
 It is worth noting that all spectra of ILRTs in our sample become bluer during the rise to maximum: such behaviour is observed in other interacting transients (e.g. SN 2023ixf, \citealt{Hiramatsu2023ixf}) and predicted by models of interacting SNe \citep{DessartBluerInteraction}, and therefore it could be a result of increasing interaction between the ejected material and the CSM. This interpretation is also supported by the increasing flux of the emission lines during the rise to maximum luminosity in all observed ILRTs (see Section \ref{LineLum_BalmerDec}). Finally, the late time spectra ($\gtrsim$100 days) collected for NGC 300 OT, AT 2019abn and AT 2019ahd do not show clear blueshift nor change in the shape of the emission lines. These are telltale signs of dust along the line of sight, and dust formation was inferred for these objects in Paper I from the NIR excess in their SED. A possible explanation for the lack of these spectroscopic signatures may be that the dust is not located along the line of sight: in this scenario, ILRTs would be intrinsically asymmetric events.
In Figure \ref{ComparisonSpectra} we display spectra of six ILRTs taken in the first weeks after maximum luminosity. Beside the four targets in our sample, SN 2008S \citep{Botticella2009} and AT 2017be \citep{CaiAT2017be} are also shown. SN 2008S is considered to be a prototype of the ILRT class thanks to its prominent Ca lines, with a peak luminosity well in line with our sample of objects. On the other hand, AT 2017be lies in the lower end of the luminosity distribution of ILRTs (see Figure 5 of Paper I).
On the right panels of Figure \ref{ComparisonSpectra} we zoom in on the most characterising features of this class, in particular the [Ca II], Ca H\&K and H$\alpha$ lines. The strength of each feature varies among the sample: for instance, AT 2019udc is characterised by a strong Ca H\&K absorption, while the [Ca II] doublet is most prominent in NGC 300 OT and SN 2008S. Albeit present, the Ca II NIR triplet becomes more prominent in all ILRTs at later phases.




\subsection{Near Infrared Spectra}


Beside the optical spectra, we gathered a total of five NIR spectra of our targets. NIR spectra of AT 2019abn were collected during the early rise, immediately after discovery. The single spectrum available for AT 2019udc was taken in a more advanced phase, three weeks after peak luminosity. Finally, the two spectra of NGC 300 OT monitor the object at late phases, hundreds of days after maximum light. As displayed in Figure \ref{NIRSpectra}, AT 2019abn and AT 2019udc appear to be quite similar: a nearly featureless continuum, without strong emission or absorption lines superimposed. 
Conversely, in the spectra collected for NGC 300 OT it is possible to identify several emission features. Most notably, these include hydrogen lines of the Paschen (Pa $\beta$,$\gamma$,$\delta$,$\epsilon$) and Brackett series (Br $\gamma$), He I ($\lambda$10830), O I ($\lambda$11287) and Fe I ($\lambda\lambda\lambda$15831,16679,16693). As mentioned in the previous section, the detection of O I $\lambda$11287 $\AA$ has been key to confirm that the optical line O I $\lambda$8446 $\AA$ is produced by the Bowen fluorescence.

\section{Line parameters}

\subsection{Line luminosity} \label{LineLum_BalmerDec}

\begin{figure*}
    \begin{center}
  \makebox[1.6\columnwidth][c]{\includegraphics[width=0.85\textwidth]
  {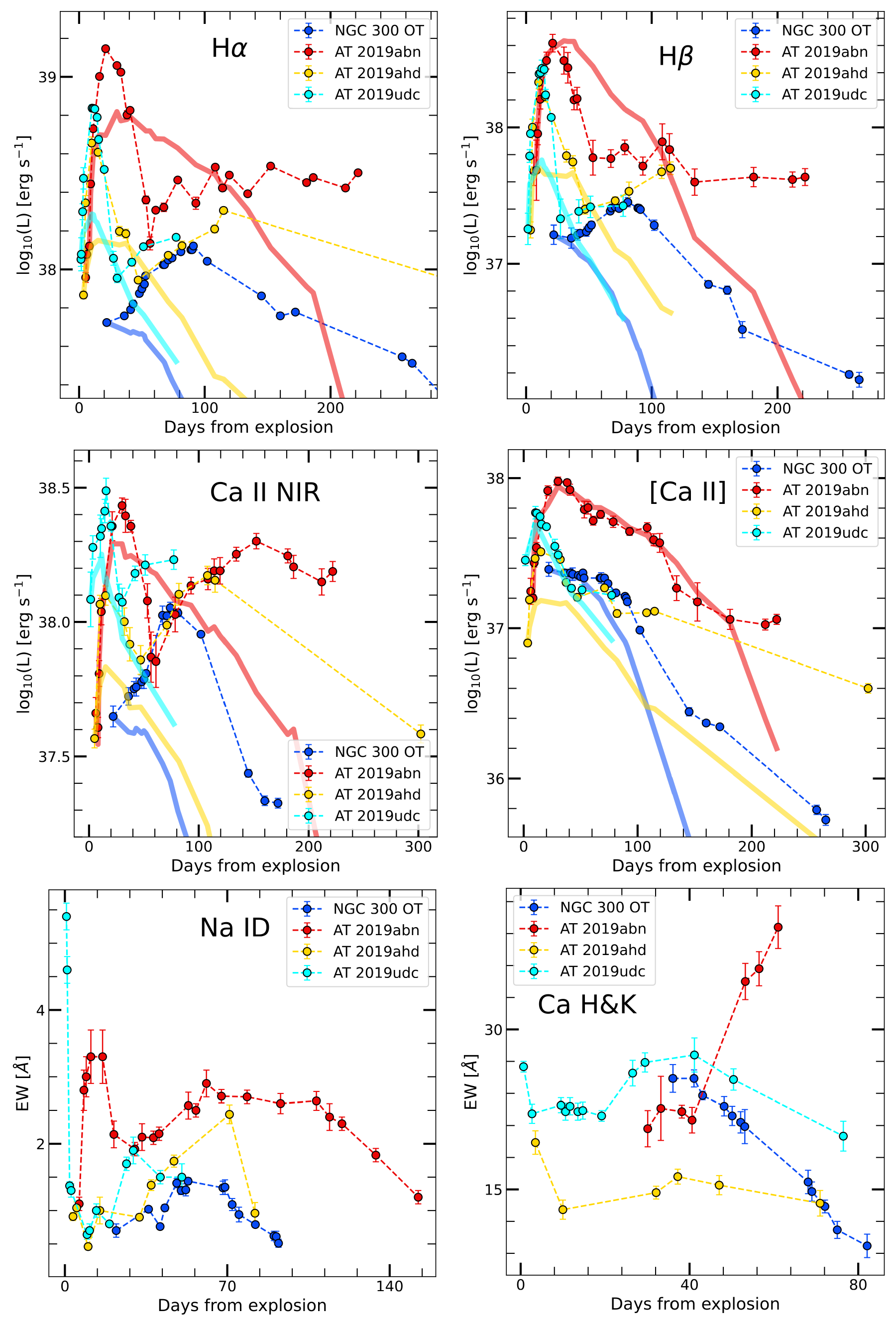}}%
  \caption{In the top four panels we show with coloured circles the luminosity evolution of the main emission features for our sample of ILRTs, namely H$\alpha$, H$\beta$, Ca II NIR triplet and [Ca II]. The continuous lines represent the level of spectral continuum in the proximity of each feature. In the bottom panels are shown the equivalent widths of the Na ID and Ca II H\&K absorption features.}
  \label{LineLuminosity}
    \end{center}
\end{figure*}


In this section we estimate the parameters of the main lines observed in our spectra. In order to avoid an additional source of systematic uncertainty, we do not assume a specific line profile (e.g. Gaussian or Lorentzian): instead we estimate the luminosity contribution of a given feature by integrating the flux over the continuum in the considered spectral region. This procedure is repeated multiple times in order to estimate the error associated with the measurement. The measured line flux is then converted into the luminosity of the line by accounting for the distance of the transient (the spectra were already corrected for extinction during the calibration procedure).
The luminosity evolution of H$\alpha$, H$\beta$, Ca II NIR triplet and [Ca II] doublet is shown in Figure \ref{LineLuminosity}. In each panel we also represent the continuum level in the proximity of each spectral feature as a solid line. For clarity, the continuum level is scaled to match the line luminosity at the first measured epoch. The continuum was estimated in the range 4725-4825 $\AA$ for H$\beta$, 6700-6850 $\AA$ for H$\alpha$ and [Ca II] and 8200-8350 $\AA$ for Ca II NIR triplet. In the same Figure, the bottom panels report the equivalent widths (EW) for Na ID and Ca II H\&K in absorption. We remark that the spectral features examined can present multiple components in high resolution (see in particular Section \ref{SECTION_highresspec}), therefore the values reported here track the behaviour of the sum of all components of each line. An additional complication is given by the fact that in the case of optically thick lines we are not sampling the whole emitting volume.

Considering the emission features, we note that the luminosity evolution of H$\alpha$ and Ca II NIR triplet is not monotonic: after an early peak, the luminosity declines sharply over the course of a few weeks. A minimum is then reached, and subsequently the lines start to strengthen again, forming a sort of broad, long lasting second peak. This trend is even more clear when compared to the evolution of the spectral continuum at the same phases, which is monotonically declining after maximum brightness. The same evolution can be detected also in H$\beta$, although less pronounced. For NGC 300 OT, only the second broad peak is visible, because the object was behind the Sun during the first phases of its evolution. This line rebrightening has already been noticed in other ILRTs and it was tentatively attributed to CSM interaction \citep{CaiAT2017be}. In this scenario, there is a collision between rapidly expanding ejecta and the slower CSM lying above. The energy released from this interaction could lead to the re-ionization of the gas above, and its subsequent recombination. The rebrightening of the hydrogen emission lines being coeval with the Ca II NIR strengthening (see Figure \ref{LinesSameObject}) suggests that these features originate in a similar environment and are produced by material undergoing similar physical processes. 
At the same time, we note that during this rebrightening of the lines there are no relevant changes in the ionisation state nor in the profile of the lines in the low resolution spectra, which are typical telltale signs of ongoing interaction. This means that either the CSM along the line of sight is optically thick enough to hide the interaction region, or that the secondary peak in line luminosity is caused by a mechanism other than CSM interaction. An alternative scenario to CSM interaction could be presence of material that is reached by the photons produced by the transient only after some time after its onset (in the order of one or two months). This material would become excited as it is overflown by photons, producing Balmer and Ca II NIR emission lines as it de-excites. Based on the onset time of the second peak (30 days for AT 2019udc, up to 55 days for AT 2019abn) and the speed of light, in this scenario the material providing the line luminosity for the secondary peak would be located at least at 8 x 10$^{16}$ cm (1.4 x 10$^{17}$ cm) from the transient. In any case, the secondary luminosity peak observed for the emission lines favours the scenario of a complex environment surrounding ILRTs, which could have been formed by a stellar wind or eruptive episodes of the progenitor star.

 Whatever the mechanism driving its evolution, the marked rebrightening shown by the Ca II NIR and H$\alpha$ lines is a characteristic commonly found in ILRTs.
 In contrast, the luminosity of the [Ca II] forbidden doublet displays an overall monotonic decline. We point out that the [Ca II] evolution appears to mimic the shape of the continuum luminosity in all four objects. This behaviour is consistent with the fact that the [Ca II] doublet originates from low density gas, farther from the dense CSM. Its evolution is therefore dictated by the incoming radiation from the transient, in particular by the radiation absorbed by the Ca H\& K lines. As previously described, the emission of the [Ca II] doublet can only happen if the medium is below the critical density. In this configuration, the hypothetical interaction between ejecta and CSM can cause the collisional excitation of Ca and H atoms in the dense CSM, which is over the critical density for the production of the [Ca II] doublet, therefore resulting in a rebrightening only for Balmer and Ca II NIR triplet lines. At the same time the slow material producing [Ca II] lines is too distant to be affected by the collision, and keeps following the evolution of the spectral continuum.



\begin{figure}
\centering
  \includegraphics[width=.95\columnwidth]{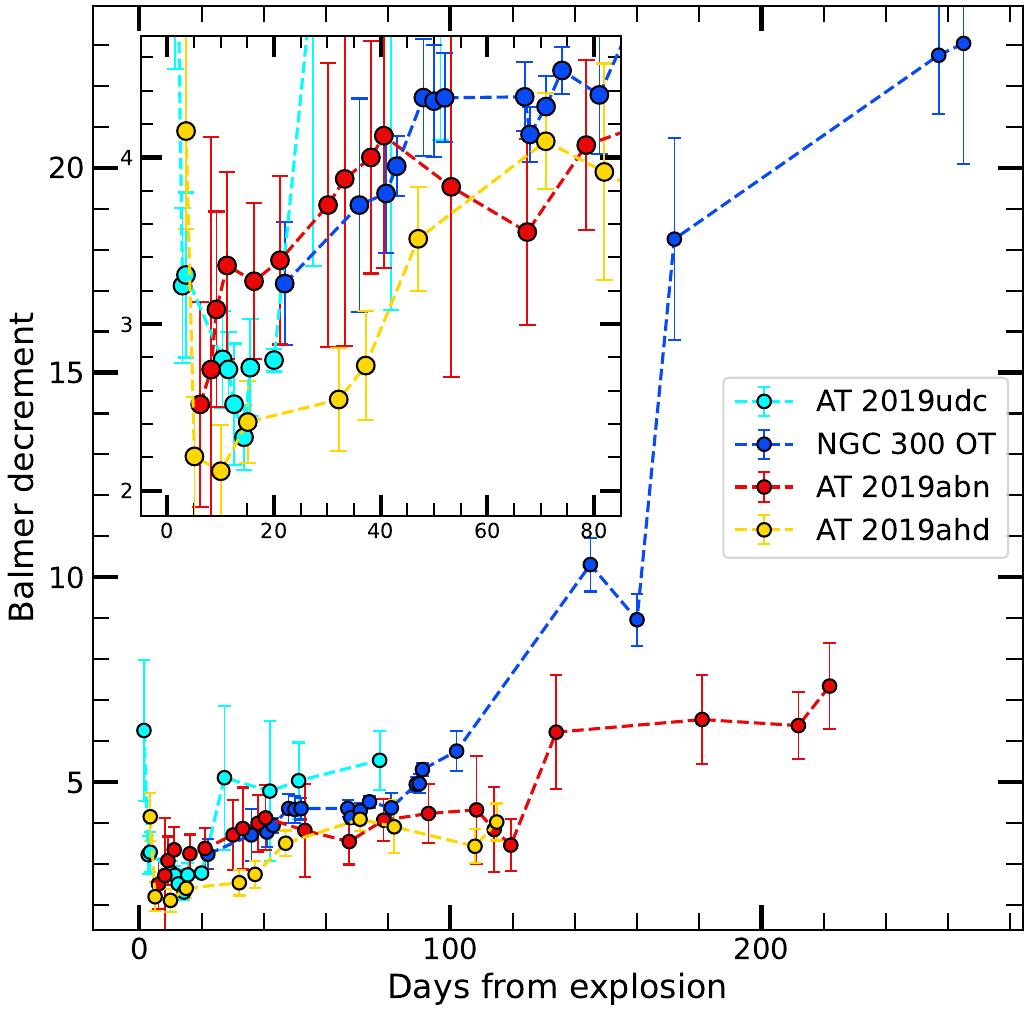}%
  \caption{Evolution of the Balmer decrement over time for our sample of ILRTs. A zoom to the early phases is shown in the small panel.}
  \label{BalmerDecrement}
\end{figure}

\begin{figure}
\centering
  \includegraphics[width=0.97\columnwidth]{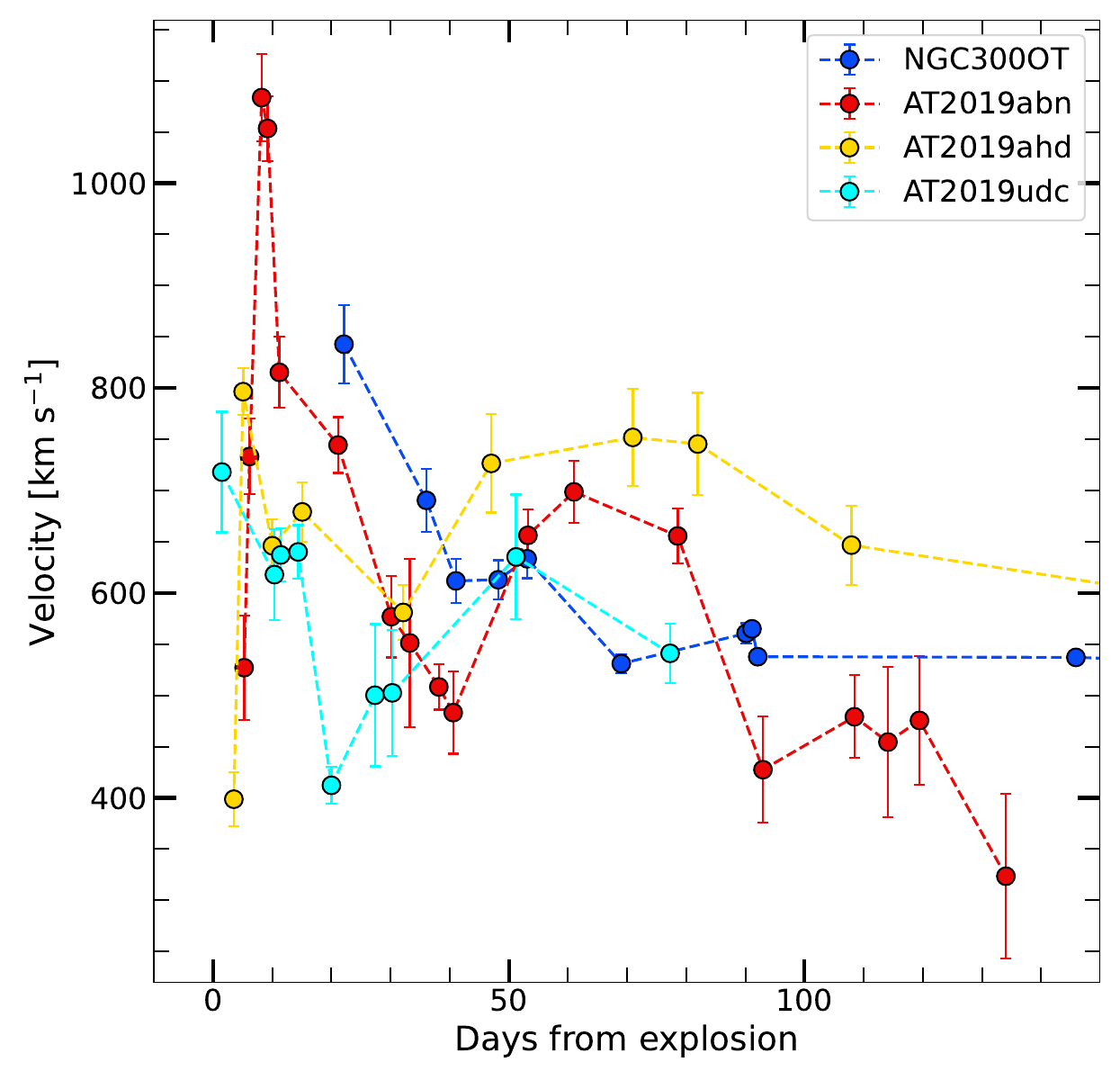}%
  \caption{FWHM velocity measured from the H$\alpha$ emission line in our sample of ILRTs during the first 150 days of their evolution.}
  \label{LineFWHMvelocity}
\end{figure}

The Na ID EW depends on the properties of the CSM, the interstellar gas along the line of sight, possible ejecta and even the viewing angle in case of non-spherically symmetric ejecta: for these reasons, its behaviour presents a large variance within the sample. AT 2019udc shows a strong Na ID feature in the first two observed epochs (EW$\sim$5 \AA), which become rapidly weaker during in the following days: a similar behaviour was observed for the ILRTs AT 2010dn and AT 2012jc \citep{Cai2021ILRTs,Byrne2023}. On the contrary, AT 2019abn initially displays a weak Na ID absorption feature (EW$\sim$1 \AA), which becomes more significant in the following weeks (up to EW$\sim$3 \AA). AT 2019ahd and NGC 300 OT present a similar behaviour, although scaled down compared to AT 2019abn. The increase and following decrease in EW of the Na ID over time can be qualitatively reproduced by considering the column density of Na I laying in front of a shock travelling through a thick CSM \citep{Byrne2023}: in this simple scenario the recombination of Na II is responsible for the initial increase in Na ID EW, while the shock travelling outward and sweeping the gas eventually causes its decline.
Finally, the EW evolution of Ca II H\&K appears to be quite similar in AT 2019ahd and AT 2019udc: an initial rapid decline followed by a somewhat broad "peak". NGC 300 OT, missing the first phases of evolution, displays a slow decline in EW that is resembling the evolution of AT 2019udc in particular. In stark contrast with the behaviour of the other three ILRTs, AT 2019abn displays Ca II H\&K absorption features that become significantly more pronounced after the first month of evolution. In principle, high resolution and multi-epoch observations of Na ID and Ca II H\&K components in ILRTs would allow us to discriminate if the evolution of these absorption features is driven by a change in the ionization state of the CSM or by the presence of clumps of matter, as prescribed by \cite{Patat2007IaCSM}. However, Na ID and Ca II H\&K absorption features often saturate in ILRTs, even despite the broadening introduced by the velocity field: this inevitably hinders the physical interpretation of these measurements.



AT 2019abn, the brightest and longest-lasting ILRT observed to date, also displays the brightest H$\alpha$ and [Ca II] features among the sample, reaching 1.6 $\times$ 10$^{39}$ erg s$^{-1}$ and 10$^{38}$ erg s$^{-1}$ at peak luminosity, respectively. Conversely, NGC 300 OT is characterised by the least luminous Balmer lines within the sample. However, beside the simple luminosity of these emission features, it is useful to investigate their evolution relative to other lines. 
For this purpose, in Figure \ref{BalmerDecrement} we track the change in value of the Balmer decrement (H$\alpha$ / H$\beta$) over time for our sample of ILRTs.
In their first spectra, both AT 2019ahd and AT 2019udc display large values for the Balmer decrement ($>$4 and $>$6, respectively), which quickly decrease down to 2.1 and 2.4 in the following two weeks. AT 2019abn and NGC 300 OT do not show this early decline, but this can be caused by the absence of very early spectra, especially for NGC 300 OT. 
With time the Balmer decrement of the whole sample grows significantly, reaching values of 7 and 23 for AT 2019abn and NGC 300 OT, respectively. 
The Balmer decrement evolution just described matches the one observed in SN 2008S. Based on the measured values of Balmer decrement and temperature at early phases, \cite{Botticella2009} constrain the electron density in the CSM to be 10$^{10}$ cm$^{-3}$ $<$ N$_{e}$ $<$ 10$^{12}$ cm$^{-3}$, according to the work performed by \cite{DrakeUlrich_BalmerDec}.
Such high electron densities are not unheard of in the context of SNe IIn: for example SN 2006gy presented an electron density in the CSM above 6$\times$10$^{10}$ cm$^{-3}$ (\citealt{Smith2006gy}).

However, these considerations are based on the assumption of case B recombination, meaning that the gas is optically thick for lines of the Lyman series and optically thin for lines of the Balmer series. Under this conditions, it has been shown that the Balmer decrement is close to $\sim$3 for a wide range of temperature and density values of astrophysical interest \citep{Osterbrock2006}. As reported in Figure \ref{BalmerDecrement}, ILRTs display a ratio H$\alpha$ / H$\beta$ which varies with time, reaching values well above the expected one of $\sim$3.
There are several possible explanations for this departure from case B recombination, such as the collisional excitation of H atoms, which would boost the H$\alpha$ luminosity, and dust formation, which would extinguish the H$\beta$ more than the H$\alpha$ \citep{DrakeUlrich_BalmerDec}. Another mechanism that can cause an increase in the Balmer decrement is an increased opacity of the lines of the Balmer series, as explained by \cite{Osterbrock2006} (Chapter 4.5).

\subsection{Width of the spectral features}

Measuring the full width at half maximum (FWHM) of the emission lines of ILRTs can be challenging: narrow features such as the [Ca II] doublet are typically unresolved in our spectra. As previously done for the line flux, we do not assume a specific fitting function, but we manually measure the FWHM of the lines. The resulting FWHM is corrected for the resolution of the instrument (typical values of R$\sim$500) and then converted into a FWHM velocity, which can be taken as an order of magnitude estimate of the speed at which gas is moving in the emitting region. This value, however, should not be treated as a precise velocity measurement: the broadening of the line profiles can be caused by the combination of Doppler shift due to bulk motion of the emitting material as well as by the electron scattering of the photons within a hot medium \citep{Fransson2014ElectronScattering}. Electron scattering is more marked at higher electron temperatures as well as higher optical depths, resulting in the appearance of extended wings \citep{HuangChevalier2018electronScatteringWing}.
In Figure \ref{LineFWHMvelocity} we display the FWHM velocities inferred from the H$\alpha$ line for our sample of ILRTs in the first 150 days of evolution. Analysing the H$\alpha$ line has the significant advantage that this feature is bright at all epochs, therefore allowing for an accurate FWHM measurement. Furthermore, especially at early times, the H$\alpha$ line is resolved in most spectra, and it is not blended with other features (as happens for the Ca II NIR triplet), making it the ideal feature for comparing the evolution of different objects with our dataset.

During the earliest phases, when the luminosity is still rising, the FWHM velocity of H$\alpha$ quickly increases in AT 2019abn and AT 2019ahd and AT 2019udc. AT 2019udc, on the other hand, displays the broadest FWHM in the first spectrum; it is impossible to confirm if this is the case also for NGC 300 OT, since we are missing the pre-maximum part of its evolution. The ILRTs in our sample reach a peak FWHM velocity between $\sim$700 and 1100 km s$^{-1}$, with AT 2019abn as a clear outlier and the other three objects clustering around v$_{FWHM}$$\sim$800 km s$^{-1}$ at peak. 
After the maximum FWHM is reached, ILRTs display a rapid decline over the course of 3-5 weeks. Afterwards, NGC 300 OT settles on an almost constant v$_{FWHM}$$\sim$550 km s$^{-1}$ for over 100 days, while the other three transients are characterised by a secondary peak in v$_{FWHM}$ after a local minimum, increasing  by $\sim$200 km s$^{-1}$ in about 5 weeks. During this phase, AT 2019ahd displays the broadest lines, with a v$_{FWHM}$ up to 750 km s$^{-1}$. Eventually the H$\alpha$ narrows once again, down to 320 km s$^{-1}$ at 135 days for AT 2019abn (and unresolved thereafter), 470 km s$^{-1}$ at 302 days for AT 2019ahd and 170 km s$^{-1}$ at 536 days for NGC 300 OT.
The behaviour of the FWHM velocity just described shows remarkable differences from the photometric features of the transients. For example, NGC 300 OT is two magnitudes fainter than AT 2019abn (see Paper I), yet its FWHM velocity is not only compatible, but oftentimes larger than the one observed for AT 2019abn.
It is important to stress that these FWHM velocities are linked to the properties of the CSM where the line is produced: as the line formation region moves within the CSM, different values of gas speed, electron temperature and optical depth mold the profile of the lines, which is reflected in the evolution of the FWHM velocity shown in Figure \ref{LineFWHMvelocity}. However, as happens in SN IIn, the low FWHM velocities measured do not exclude the presence of fast ejecta lying below the CSM (e.g. \citealt{1998S}).



\begin{figure*}
   \begin{center}
  \makebox[2.0\columnwidth][c]{\includegraphics[width=0.91\textwidth]{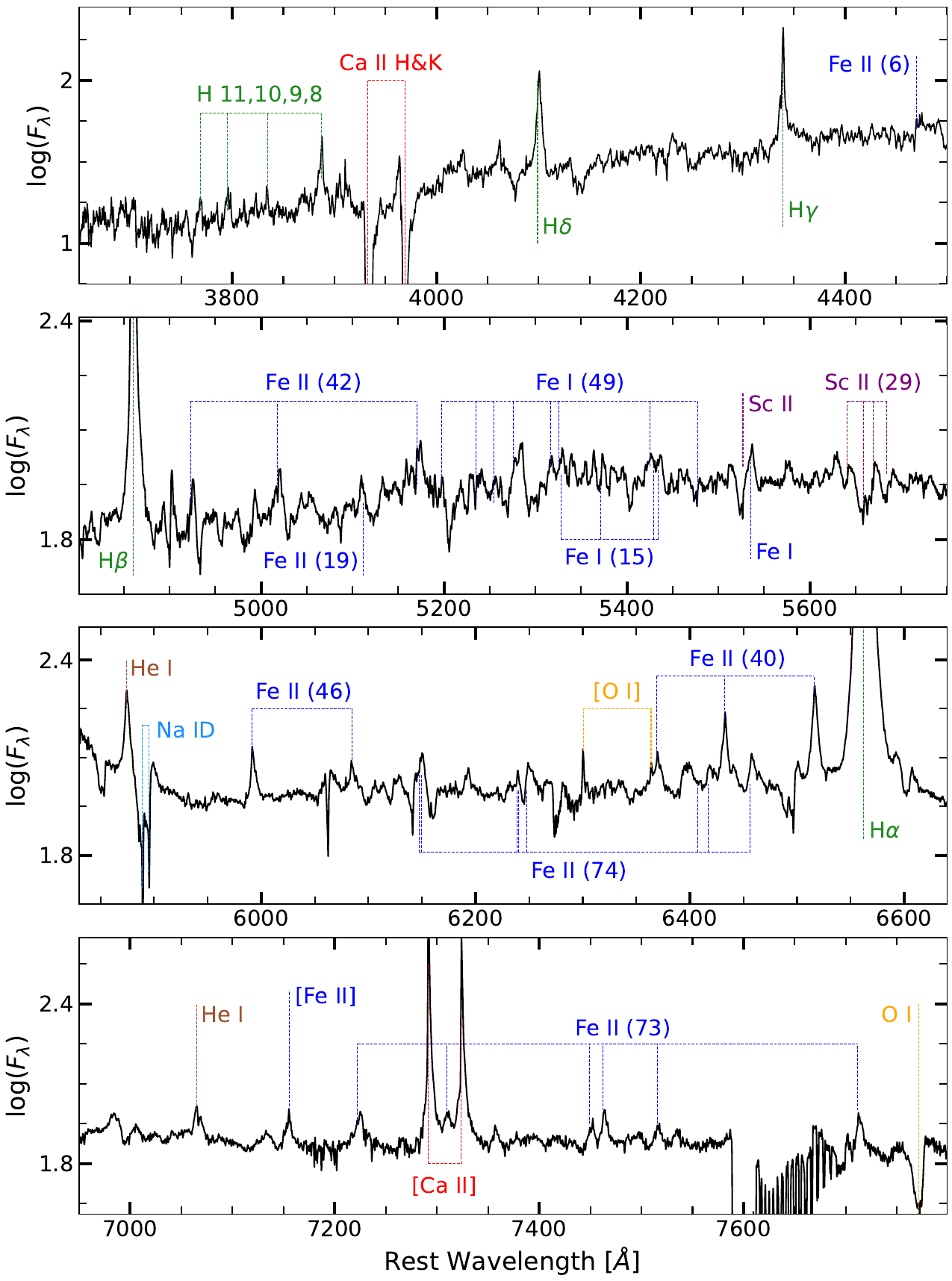}}%
  \caption{Identification of the main spectral lines found in the UVES spectrum of NGC 300 OT taken on 2008-07-02. Lines originating from different elements are highlighted in different colours. The spectrum has been rebinned for clarity.}
  \label{LineIdentification}
   \end{center}
\end{figure*}

\begin{figure*}
 \begin{center}
  \makebox[1.6\columnwidth][c]{\includegraphics[width=0.8\textwidth]
  {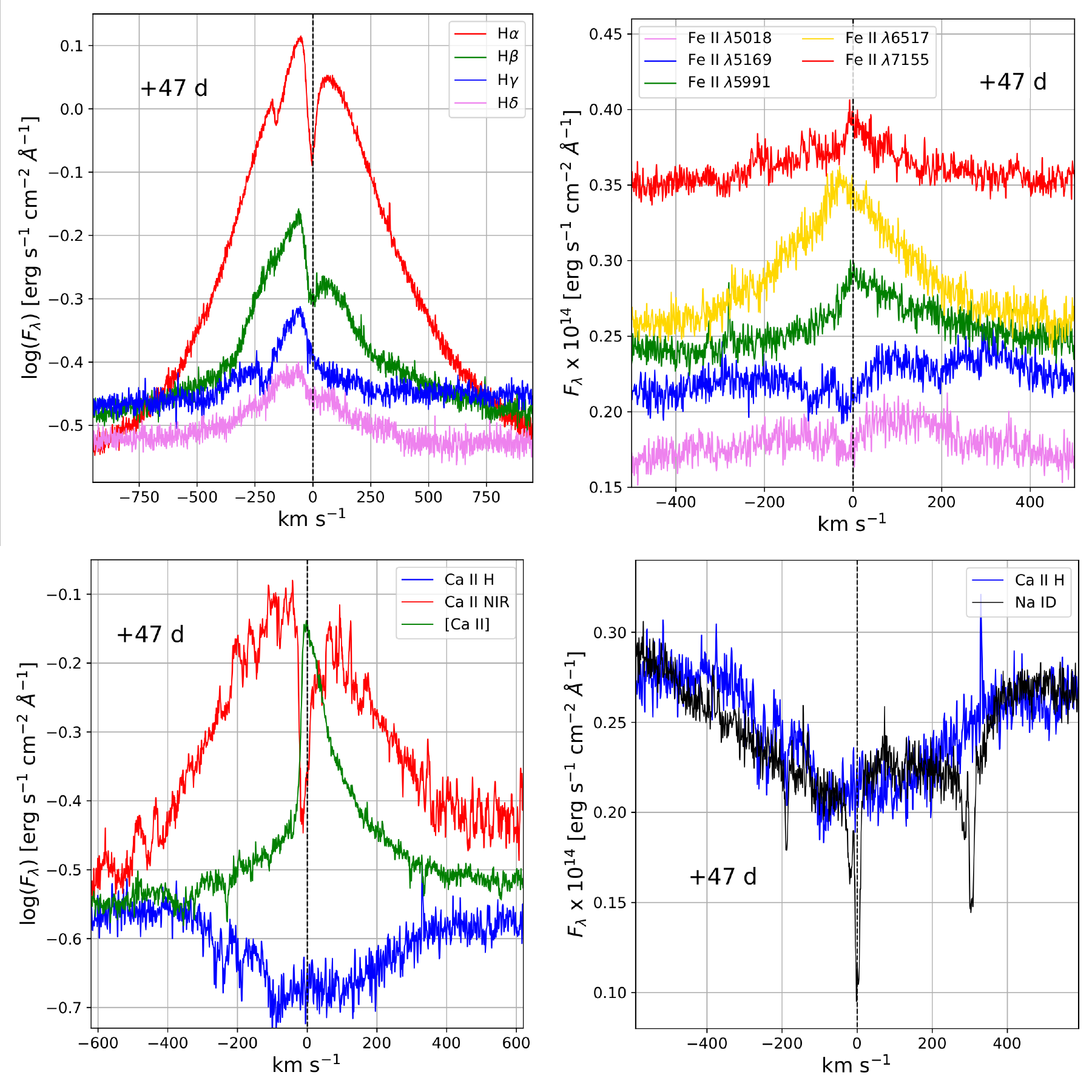}}%
  \caption{The most prominent features in the high resolution spectrum of NGC 300 OT displayed in the velocity space. In the top left panel the lines of the Balmer series d are shown. In the top right panel we display the different profiles of Fe II lines. In the bottom left the Ca lines visible in the UVES spectrum are compared: Ca II NIR, [Ca II] and Ca II H. Finally, in the bottom right the complex Na ID absorption features are shown, compared with the Ca II H absorption at the same epoch. The spectra have been scaled in flux for clarity.}
  \label{LineHDprofiles}
   \end{center}
\end{figure*}


\section{High resolution spectroscopy of NGC 300 OT} \label{SECTION_highresspec}

NGC 300 OT is the closest ILRT observed (see \citealt{Cai2021ILRTs} and \citealt{Karambelkar2023} for a review of known ILRTs). Thanks to the proximity of its host galaxy (1.92$\pm$0.14 Mpc), it was possible to obtain a high resolution optical spectrum of the transient on 2008-07-02 with the Very Large Telescope (VLT) equipped with UVES, yielding a resolution between 0.12 and 0.36 $\AA$. In Figure \ref{LineIdentification} the identification of the main spectral lines on this high resolution spectrum is shown. The measurements pertaining to the most prominent features are reported in Table \ref{LineID}, in particular their observed central wavelength, their FWHM and their estimated flux.
Apart from the H, Ca, O and Na lines discussed in the previous sections, easily recognizable also in the low resolution spectra, in this high resolution and high S/N we find a number of metal lines, in particular Fe I ($\lambda\lambda\lambda\lambda$ 5535,8688,8757,8824), Fe II (multiplets 40,42,46,49,73,74 and forbidden line $\lambda$ 7155), Mg I ($\lambda$8806) and Sc II lines (Moore multiplet 29). Additionally, we identify [O I] ($\lambda\lambda$ 6300,6364), faint Ca II absorption doublet ($\lambda\lambda$8912,8927) and He I ($\lambda\lambda\lambda$ 5876,6678,7065). All these features are narrow (FWHM $\lesssim$ 10 \AA) but resolved, therefore originating within the opaque CSM or above it. 
Thanks to the high resolution spectrum, it is possible to accurately compare the FWHM velocities associated with different emission features. The broadest feature observed at this epoch is the H$\alpha$ emission line, with a FWHM velocity of 520 km s$^{-1}$. The Ca II NIR line $\lambda$ 8498 is marginally narrower, with a FWHM velocity of 440 km s$^{-1}$: the fact that H$\alpha$ and Ca II NIR have compatible shapes in velocity space corroborates the scenario outlined in the previous section, where both these features are generated in neighbouring environments. The [Ca II] forbidden doublet, instead, is characterised by a FWHM of 80 km s$^{-1}$, pinpointing its origin in a separated region, likely characterized by lower gas velocity, electron temperature and optical depth compared to the region where permitted lines are formed. The [O I] forbidden lines are comparable in FWHM and shape with the [Ca II] doublet.
In general, the range of FWHM velocity displayed by the lines (Table \ref{LineID}) suggests a stratification in the emission lines of different ions, and even within the emission lines of the same ions. 


\begin{table}

\begin{adjustbox}{tabular= ccccc, center}
 \\ \hline
 Line & $\lambda_{0}$ & $\lambda_{m}$  & FWHM & Flux \\ 
  & [\AA] & [\AA] & [\AA] & [$10^{-15}$ erg s$^{-1}$cm$^{-2}$]\\ 
 \hline
 H$\alpha$ & 6562.8 & 6562.5 & 11.6 & 235.8\\
 H$\beta$ & 4861.3 & 4860.3 & 5.8 & 54.2\\
 H$\gamma$ & 4340.5 & 4339.4 & 2.4$^{*}$ & 17.0$^{*}$\\
 H$\delta$ & 4101.7 & 4100.7 & 5.0 & 13.5\\
 Ca II NIR & 8498.0 & 8495.9 & 12.6 & 97.4 \\
 \big[Ca II\big] & 7291.5 & 7292.0 & 1.9 & 21.7 \\
 \big[Ca II\big] & 7323.9 & 7323.4 & 2.0 & 15.6 \\
 Ca II H & 3933.7 & 3933.9 & 6.4 & - \\
 Ca II K & 3968.4 & 3968.9 & 4.7 & - \\
 Fe II (40) & 6516.1 & 6516.7 & 6.0 & 10.4 \\
 Fe II (40) & 6432.7 & 6432.8 & 4.9 & 6.7 \\
 Fe II (40) & 6369.5 & 6368.7 & 6.2 & 4.6 \\
 Fe II (42) & 5169.0 & 5170.7 & 7.0 & 6.6 \\
 Fe II (42) & 5018.4 & 5018.6 & 8.0 & 5.4 \\
 Fe II (42) & 4923.9 & 4925.2 & 5.2 & 2.9 \\
 Fe II (46) & 5991.4 & 5992.5 & 4.5 & 4.2 \\
 Fe II (46) & 6084.1 & 6084.4 & 6.2 & 3.3 \\
 He I & 5875.6 & 5874.9 & 5.0 & 7.6 \\
 He I & 7065.2 & 7065.7 & 8.2 & 3.8 \\
 \big[O I\big] & 6300.3 & 6300.4 & 1.6 & 1.4 \\
 \big[O I\big] & 6363.8 & 6363.9 & 1.6 & 0.6 \\
 O I & 7771.9 & 7770.3 & 9.9 & - \\
 Na I & 5889.9 & 5889.5 & 5.5 & - \\
 Na I & 5895.9 & 5895.5 & 6.5 & - \\
 \hline
\end{adjustbox}
\caption{Parameters of the main lines identified in the UVES spectrum of NGC 300 OT taken on 2008-07-02. The columns report the identified line, its rest wavelength, the FWHM, the FWHM velocity and flux obtained as detailed in the text. Errors on the flux and FWHM measurements are of the order of 10\%. The spectrum was redshift and reddening corrected with the values reported in Table \ref{tab_Dist_Redd}. \\
$^{*}$Affected by absorption features, as detailed in Section \ref{SECTION_highresspec}.}
\label{LineID}
\end{table}

Spectral features that appear simple in low resolution may actually hide a more complex structure, which can only be unveiled through high resolution observations. The shape of such lines carries relevant information on the geometry of the emitting region. In Figure \ref{LineHDprofiles} we show the profiles of the lines of the Balmer series, of some prominent Fe II lines, of the Ca features (NIR $\lambda$8498, [Ca II] $\lambda$ 7291 and Ca II K absorption) and the Na ID absorption doublet. The H$\alpha$ line in the UVES spectrum shows a composite profile: superimposed to the emission line are two narrow absorption components. The most prominent absorption component has a FWHM of 30 km s$^{-1}$ and is only 15 km s$^{-1}$ redward of the rest wavelength of H$\alpha$. The second absorption component is even narrower, with a FWHM close to the spectral resolution, and it is situated 140 km s$^{-1}$ blueward of the emission peak. In the top left panel of  Figure \ref{LineHDprofiles}, the most prominent and redward of the two absorption features can be seen in other coeval lines of the Balmer series, although with lower S/N. The second, more shallow and blueward absorption does not have a counterpart in the other H lines, and therefore it could result from a transition of a different element. The H$\gamma$ line presents a unique shape, which deviates from the trend of the other Balmer lines, showing in particular a marked absorption in the blue wing which also results in a narrower FWHM. The same behaviour for the H$\gamma$ line has been observed in the yellow supergiant IRC+10420, where it has been explained as the result of absorption components found in stellar atmosphere models (Figure 6 in \citealt{Klochkova_Hgamma_IRC+10420}). \\
In the top right panel of Figure \ref{LineHDprofiles} it is possible to study the different profiles of the Fe II lines. Despite originating from the same atomic species, some lines display a dominant emission component, with only traces of absorption at around the rest wavelength (e.g. $\lambda$6517), while other lines display a more complicated profile, with marked absorption features and a redshifted emission component, compatible with a P-Cygni profile (e.g. $\lambda\lambda$5018,5169). The Fe II lines that display a stronger absorption, such as the multiplet 42, are characterized by a larger oscillation strength, which favours self absorption. Additionally, the spatial stratification of the emission lines may further enhance this effect.

The analysis of the Ca II lines (bottom left panel of Figure \ref{LineHDprofiles}) offers valuable insight on the geometry of the line forming region.
The Ca II NIR line $\lambda$8498 displays, as said, a prominent absorption component with a FWHM of 30 km s$^{-1}$ situated just 10 km s$^{-1}$ blueward of the rest wavelength position of this line. 
On the other hand, the [Ca II] doublet presents a very different line profile: the line is markedly asymmetric, with the blue wing almost completely missing. \cite{Berger2009} tentatively attribute this behaviour to self absorption, in a scenario where the [Ca II] doublet originates in an low density, inflowing gas. However, \cite{Hump2011} state that self absorption is quite rare for a forbidden transition such as [Ca II] (although not impossible, as shown for example by \citealt{ElenaRevisitingV1309Sco}): the peculiar line profile is postulated to be formed through electron scattering in an expanding envelope, which gives rise to the extended red wing observed.
We expand on this interpretation, based on the similarity between the shape of the absorption feature superimposed to the Ca II ($\lambda$8498) emission line and the [Ca II] doublet: a steep blue edge and an extended red wing (lower left panel of Figure \ref{LineHDprofiles}). We deduce that both features are produced in the same region, in which the gas density must be low to allow the production of the [Ca II] doublet. 
The velocity field of the gas in this region molds the shape of both spectral features. We speculate that the observed asymmetry reflects the dynamics of a stellar wind which reached the terminal velocity. The blueward sharp cutoff would indicate the terminal wind velocity.
By extension, all the observed narrow absorptions superimposed to the emission lines (Balmer lines, Ca II NIR triplet, He I, Fe II and O I) can be traced back to this low density, pre-existing wind produced by the progenitor star. With the redshift we adopted for NGC 300 OT, the blue cutoff for the [Ca II] is located at $\sim$-30 km s$^{-1}$, which provides an estimate of the terminal wind velocity in the scenario just described. Such value is well in line with the wind velocities observed for AGB stars (e.g. \citealt{AGBwindVel}).
On the other hand, the red wing of the [Ca II] extends over 200 km s$^{-1}$ redward: this could be the result of photons scattered into the line of sight by interaction with free electrons. Indeed, a similar line profile has been obtained by simulating the scattering of an emission line within a stellar wind (Figure 10 in \citealt{HuangChevalier2018electronScatteringWing} and Figure 4 in \citealt{JerkstrandNebular}). \\
In the Na ID absorption profile (bottom right of Figure \ref{LineHDprofiles}) it is possible to differentiate multiple components: a narrow, barely resolved (v$_{FWHM}$$<$10 km s$^{-1}$) component associated with the Milky Way gas (which is unchanging with time, according to \citealt{Berger2009}), a second narrow component associated with the host galaxy, NGC 300, and a broader absorption (FWHM velocity $\sim$200 km s$^{-1}$ for each component of the doublet) originating from the gas expanding around the transient (for a more detailed discussion of Na ID in ILRTs, see \citealt{Byrne2023}).
Lastly, the Ca II H line presents a FWHM velocity of 420 km, ruling out an interstellar origin in favour of production within the more turbulent CSM surrounding the transient. The other line of the doublet, Ca II K, is contaminated by the H$\epsilon$ emission line, resulting in a misleading asymmetric profile with a short blue wing and an extended red wing.


\begin{figure*}[hbtp!]
  \makebox[\textwidth][c]{\includegraphics[width=1.00\textwidth]{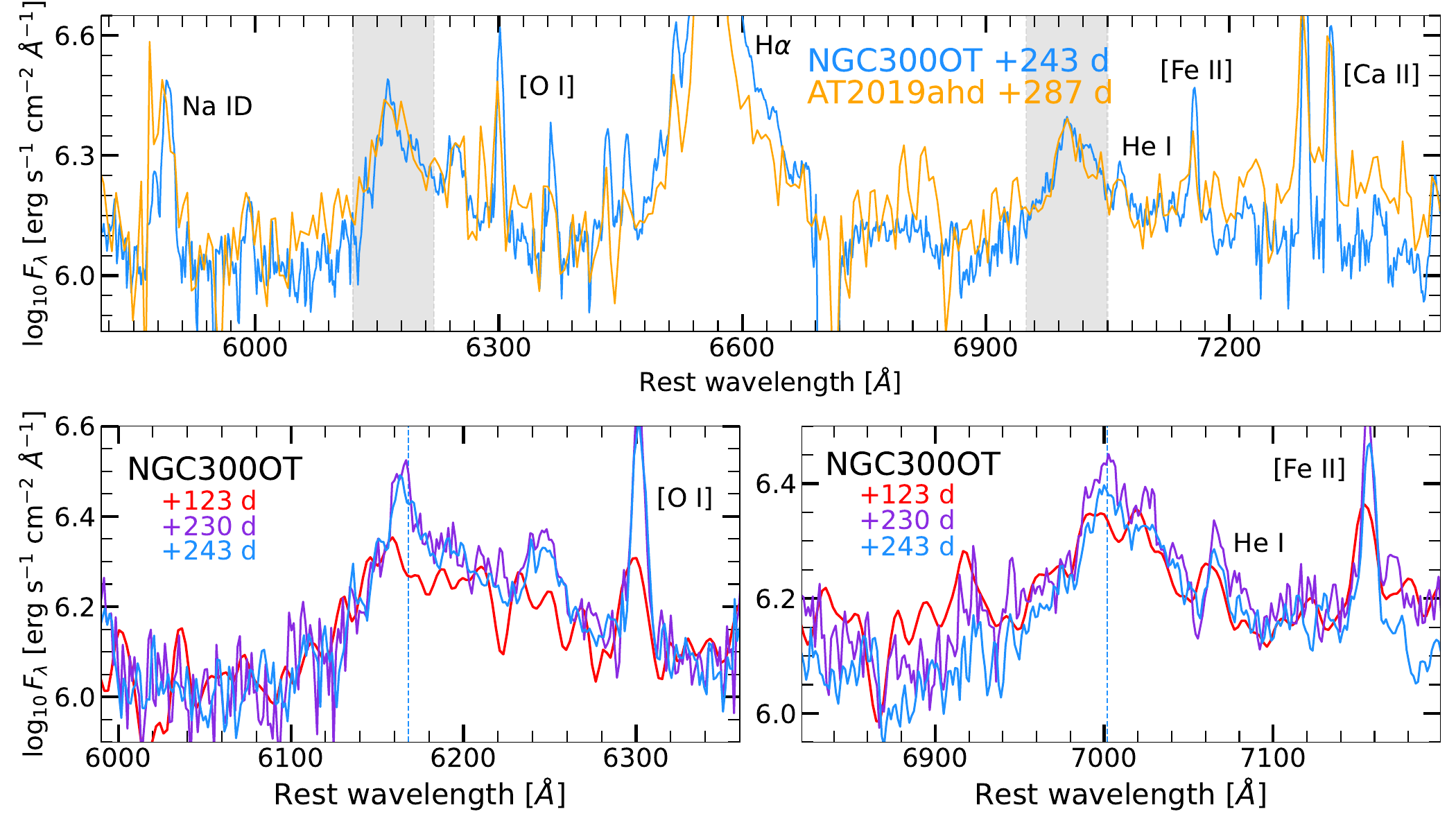}}%
  \caption{Late time spectra of NGC 300 OT and AT 2019ahd. A shift in flux has been applied to superimpose the two spectra. The broad features described in the text are highlighted with gray shaded regions. In the lower panels is shown the temporal evolution of these broad features in NGC 300 OT, starting from their first detection. Dashed vertical lines mark the position of their peak estimated through a gaussian fit.}
  \label{Late2019ahdNGC300OTspec}
\vspace{0.3cm}
  \makebox[\textwidth][c]{\includegraphics[width=1.00\textwidth]{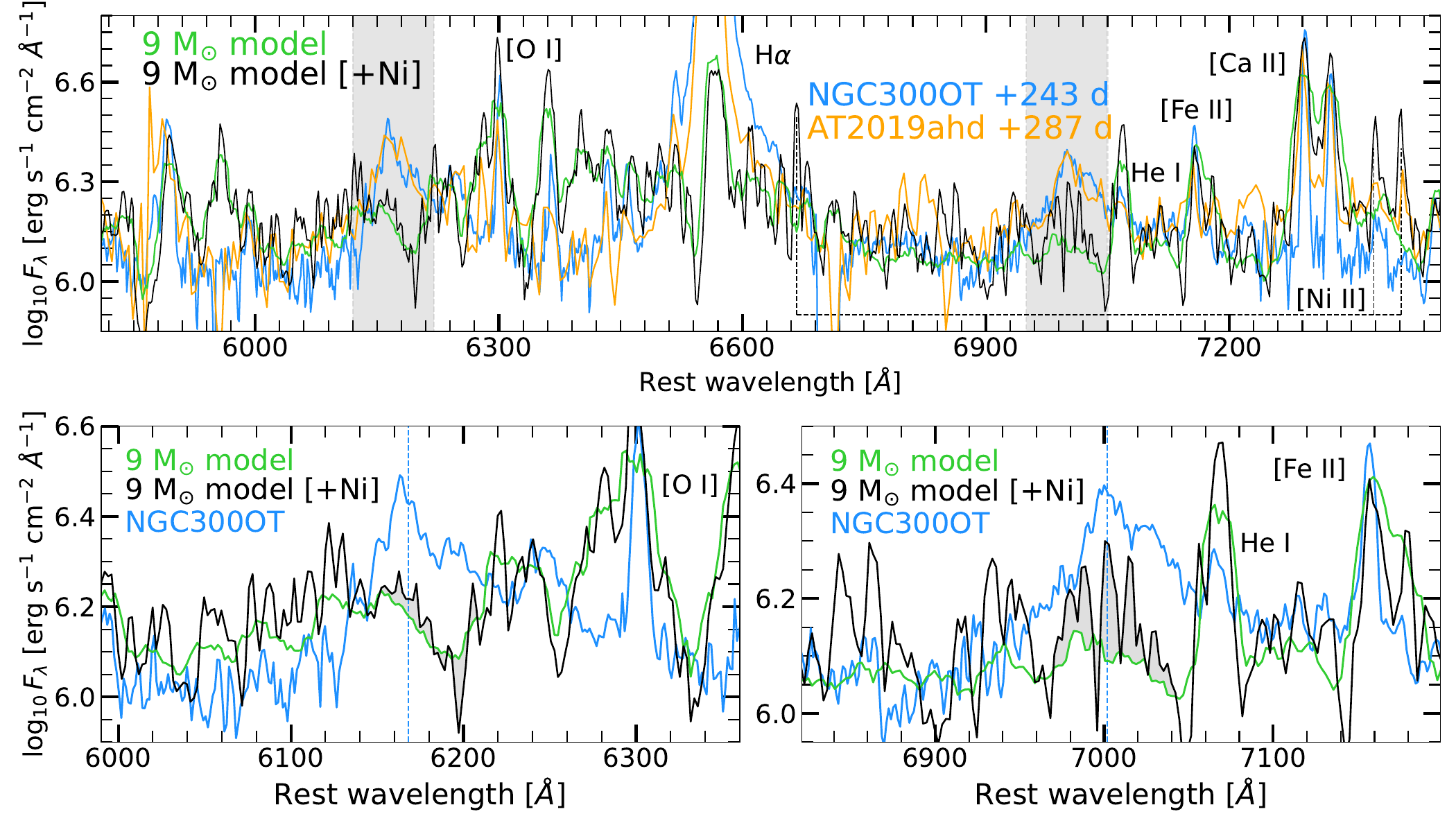}}%
  \caption{Late time spectra of NGC 300 OT compared with models of nebular spectra taken from \protect \cite{JerkstrandModels}. Forbidden [Ni II] lines expected in Ni-rich ejecta are marked with dashed lines. In the bottom panels is shown a zoom on the region where broad lines are observed in ILRTs. The shaded gray area in the bottom panels highlights the flux difference in the two models.}
  \label{SpecModNGC300}
\end{figure*}




\section{Broad features in late time spectra}

Late time spectra of ILRTs display some intriguing features. Both in NGC 300 OT and AT 2019ahd, two broad features start appearing at 6168 $\AA$ and 7002 $\AA$ a few months after peak luminosity (the position of the peak was estimated through a Gaussian fit). Notably, so far such broad lines were never detected in any spectrum of ILRTs. In Figure \ref{Late2019ahdNGC300OTspec} we show the late time spectra of NGC 300 OT and AT 2019ahd superimposed, in order to highlight the similarities between the two, in particular regarding these broad features. The red wing of the bluer feature (at 6168 $\AA$) partially overlaps with the narrow [O I] $\lambda$6300 line, while another emission feature can be detected at $\lambda$6247. On the redder feature (at 7002 $\AA$) there is a minor contamination from He I $\lambda$7065. It is challenging to exclude the presence of absorption components, similar to the ones detected in the H and Ca II NIR lines, given the rather low S/N. Both broad features present a FWHM velocity of $\sim$2500 km s$^{-1}$.
The lower panels of Figure \ref{Late2019ahdNGC300OTspec} show their evolution over time in NGC 300 OT, from their first detection at +123 d up to +243 d. Their development is remarkably slow, as the features appear barely different after 120 days. As for AT 2019ahd, we only detect the broad emission components in a single spectrum, therefore it is not possible to study their evolution in time.
Given the rather symmetrical distribution of these broad features around the H$\alpha$ line, in principle they could be explained through H$\alpha$ emission from two fast jets of matter moving at $\sim$20000 km s$^{-1}$ (based on the displacement of their peak with respect to the H$\alpha$ rest wavelength), one approaching and one receding along the line of sight. However, since the features appear to be almost identical in both AT 2019ahd and NGC 300 OT, it is extremely unlikely to have observed two jets oriented exactly in the same direction relative to the Earth, and therefore we disfavour this scenario.

\begin{figure*}[hbtp!]
  \makebox[\textwidth][c]{\includegraphics[width=1.0\textwidth]{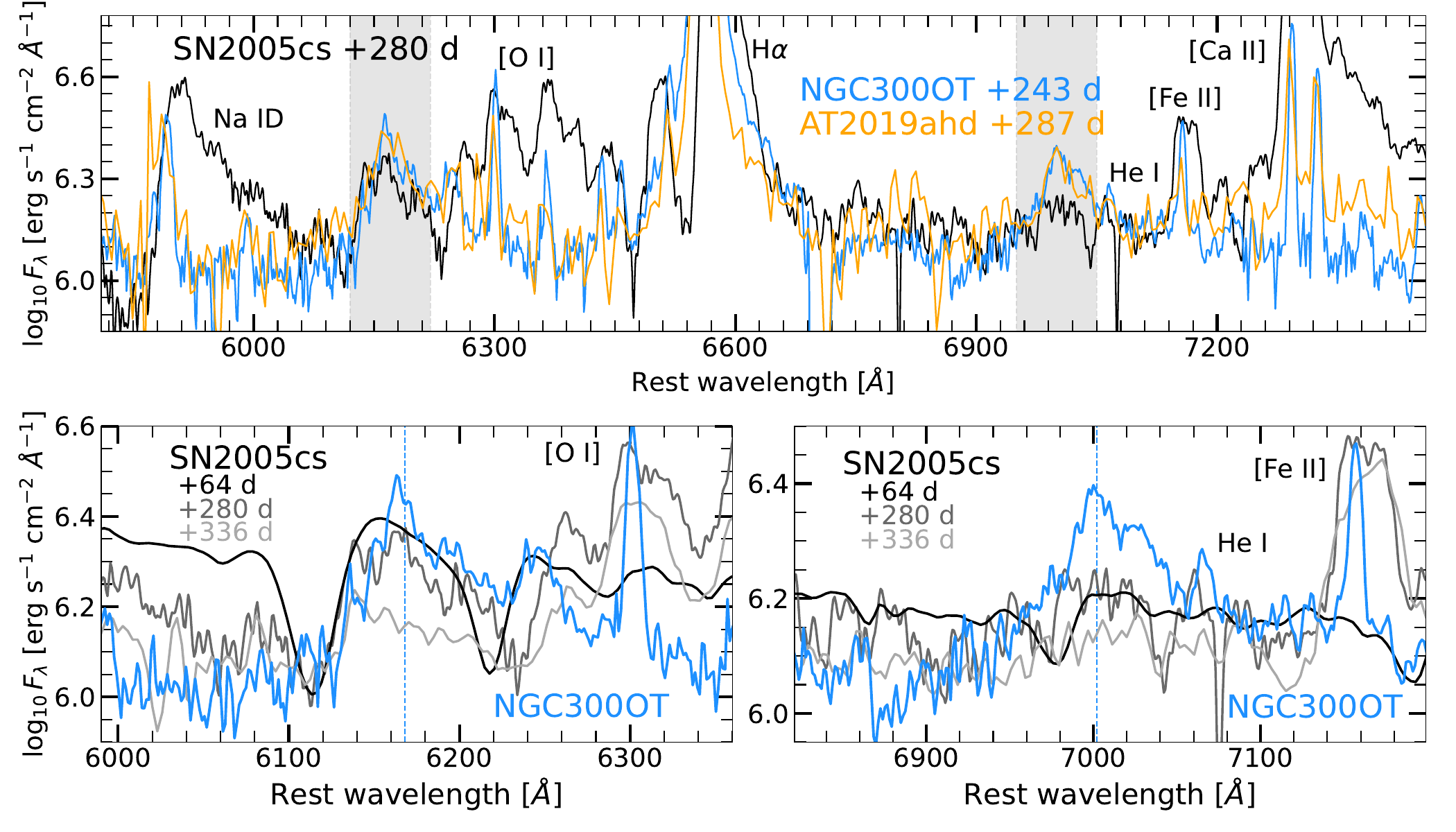}}%
  \caption{Late time spectra of NGC 300 OT and AT 2019ahd compared with SN 2005cs. The lower panels show the temporal evolution of SN 2005cs in the considered wavelength ranges compared with NGC 300 OT.}
  \label{Late_IIP_ILRT}
\vspace{0.8cm}
  \makebox[\textwidth][c]{\includegraphics[width=1.0\textwidth]{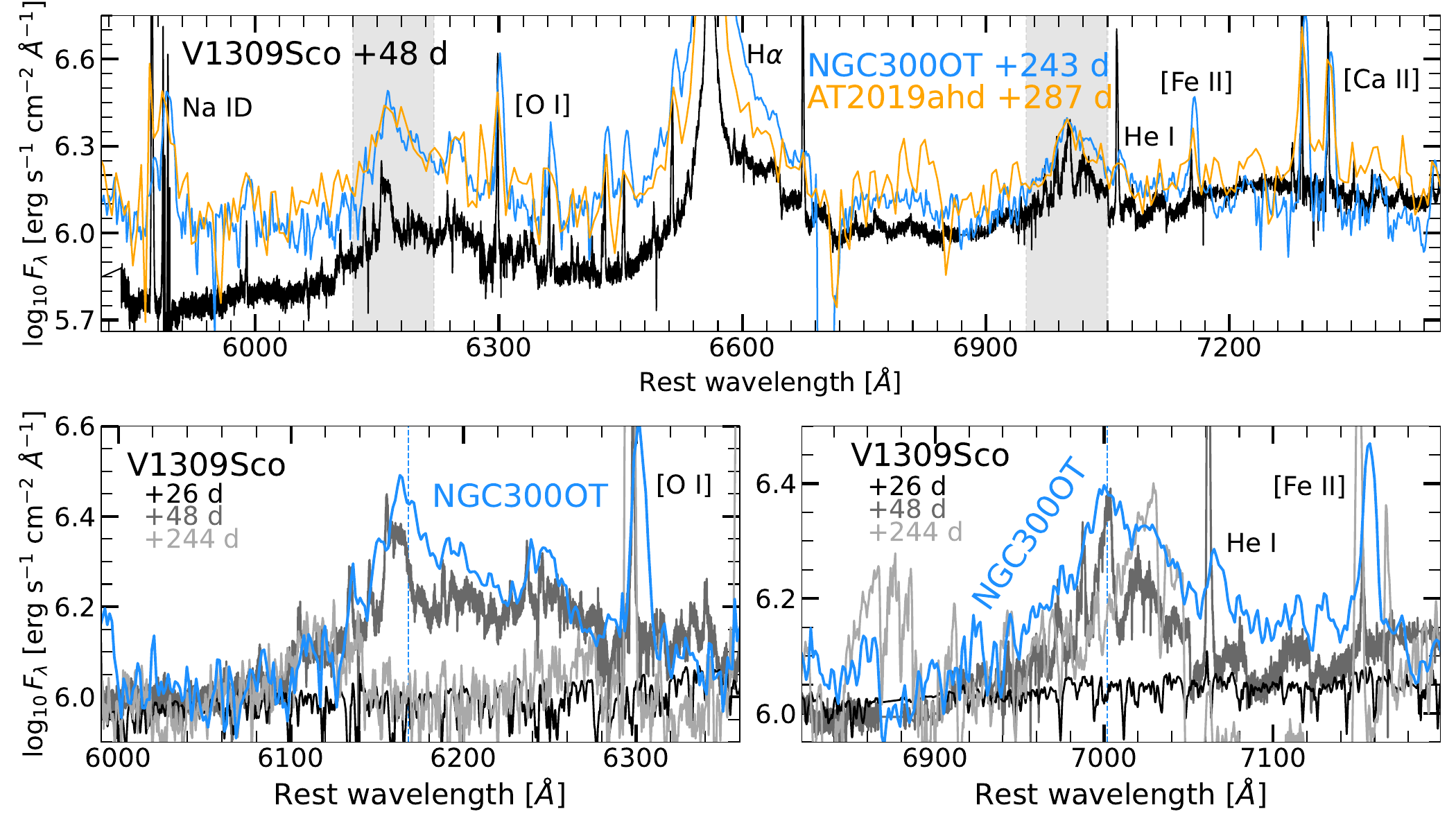}}%
  \caption{Late time spectra of NGC 300 OT and AT 2019ahd compared with V1309Sco. The lower panels show the temporal evolution of V1309Sco in the considered wavelength ranges compared with NGC 300 OT.}
  \label{Late_LRN_ILRT}
\end{figure*}


In the context of a terminal, explosive event, it is tempting to associate these features with the products of nucleosynthesis emerging from the slow moving CSM that was obscuring our line of sight during the photospheric phase. However, these broad features are challenging to identify, as they do not match the common nebular lines found in SNe (e.g. \citealt{BenettiWheelerNebular}).
Few atomic species can account for both the emission lines: according to the tabulated values reported by \cite{MooreTables1945}, a candidate could be Ni I $\lambda\lambda 6177,7002$, which would corroborate the scenario in which explosive nucleosynthesis took place during the ILRT events.
In order to test this hypothesis, in Figure \ref{SpecModNGC300} we compare the late time spectra of NGC 300 OT and AT 2019ahd with the models of nebular spectra presented by \cite{JerkstrandModels}. The synthetic spectra have been scaled down for clarity. These models aim to reproduce the late time spectra of SNe arising from the core collapse of a 9 M$_{\odot}$ star (at zero age main sequence). The first synthetic nebular spectrum considered is relative to 300 days after explosion and has a solar abundance of elements. A high Ni content is expected to be an indicator of an ECSN event \citep{Wanajo}: for this reason we consider a second synthetic spectrum (again provided by \citealt{JerkstrandModels}), at the same evolutionary stage, in which the ratio Ni/Fe is 30 times the solar value. In the lower panels of Figure \ref{SpecModNGC300} we display a zoomed in view of the broad lines found in NGC 300 OT, together with the same spectral region of the synthetic nebular spectra. In the region of the feature peaking at $\sim$7002 \AA, there is a noticeable contribution in flux that is present in the Ni rich spectrum. 
However, the broad emission component at 6168 $\AA$ is not well reproduced by either synthetic spectrum, and an enhanced abundance of Ni does not seem to significantly influence this spectral region. Furthermore, a notable shortcoming of this interpretation is the lack of prominent [Ni II] lines ($\lambda$$\lambda$$\lambda$6667,7378,7412, marked with dashed lines in the upper panel of Figure \ref{SpecModNGC300}) in our ILRTs spectra, which would instead be expected in the nebular spectra of a Ni-rich explosion \citep{JerkstrandModels}. Having considered these elements, we deem unlikely that the broad features observed originate from the nebular emission of fast, Ni-rich ejecta.



We now perform comparisons with empirical data, starting with the well studied SN 2005cs, a low luminosity SN IIP \citep{Pasto2005csII}. In the upper panel of Figure \ref{Late_IIP_ILRT} we plot the nebular spectrum of SN 2005cs, taken 280 days after explosion, together with the late time spectra of NGC 300 OT and AT 2019ahd. We observe a promising match for the broad feature at 6168 $\AA$: interestingly, during the photospheric phase of SN 2005cs there are two absorption components compatible with P-Cygni profiles in that region (bottom left panel of Figure \ref{Late_IIP_ILRT}). The broad emission components could therefore originate from the blend of multiple Fe II (and possibly Fe I) lines, which form a P-Cygni profile during the photospheric phase of SN 2005cs (+64 d), but then turn to pure emission when the continuum fades (+280 d). Subsequently, almost one year after explosion (+336 d), the broad emission component appears weakened. On the other hand, SN 2005cs does not present strong emission around the region around 7002 $\AA$, as shown in the bottom right panel of Figure \ref{Late_IIP_ILRT}, although there could arguably be a feeble P-Cygni profile in this region during the photospheric phase.

As a final comparison, we present the red nova V1309 Scorpii (hereby V1309 Sco, \citealt{MasonEV1309Scorpii}), which is interpreted as the result of a merger of a contact binary system \citep{TylendaV1309Sco,KaminskiV1309Sco}. The spectrum taken 48 days after the start of the outburst shows remarkable similarities with the late time spectra of our ILRTs (Figure \ref{Late_LRN_ILRT}). Thanks to the high resolution provided by the VLT equipped with UVES, it is possible to distinguish multiple narrow features in the spectrum of V1309 Sco which corroborate the scenario in which the broad emission components originate from the blend of multiple narrow lines. Fe I and II lines are likely candidates, although pinpointing the exact transitions involved is quite challenging. As shown in the bottom panels of Figure \ref{Late_LRN_ILRT}, the insurgence of this forest of emission lines is sudden, since there is no hint of emission over the continuum in the regions around 6168 $\AA$ and 7002 $\AA$ in the spectrum of V1309 Sco taken 26 days after the outburst. 
Based on this comparison, we favour the interpretation that the broad emission lines observed in late time spectra of ILRTs are blends of multiple narrow (possibly Fe) lines, as observed for V1309 Sco.


This newly discovered resemblance between the spectra of V1309 Sco, AT 2019ahd and NGC 300 OT shows that similar chemistry and physical conditions of the emitting regions can be found in red novae and ILRTs. However, while there are some objects showing hybrid properties (e.g. AT 2022fnm \citealt{2022fnm}), this does not mean that ILRTs and LRNe can be easily blended into one class.
First of all, the timescales over which V1309 Sco and the ILRTs develop these broad emission features are quite different. This could be linked to the different scale of the phenomena: the peak optical luminosity of V1309 Sco is about $\sim$150 times lower than the peak luminosity of NGC 300 OT and AT 2019ahd, and it is characterised by a faster decline rate, dimming by 3 magnitudes in 50 days. 
Furthermore, late time spectra of red novae such as V1309 Sco are characterised by a red continuum eroded by the absorption of countless molecular bands, most notably TiO, VO, CO and H$_{2}$O (e.g. \citealt{GoranskijMolecularBands}).
 On the contrary, ILRTs never display the absorption features typical of molecular bands. This difference is made evident by comparing the spectrum of NGC 300 OT taken 514 days after maximum with X-Shooter with the spectrum of a red nova that already developed molecular band absorptions. Even at 514 days after maximum luminosity, the surviving continuum of NGC 300 OT does not show signs of molecular bands, which are instead prominent in the case of V1309 Sco, 244 days after the outburst (see Figure \ref{MolecularBandsNIR} in Appendix).

\section{On the nature of ILRTs}

The physical mechanism driving ILRTs has been a puzzle since the identification of this class of objects, 15 years ago. We quickly outline the main observed features of these transients before reviewing some scenarios that could explain their origin:

\begin{itemize}
  \item[$\bullet$] ILRTs present a single peaked, monotonically declining light curve. The total radiated energy spans from few 10$^{47}$ to few 10$^{48}$ erg (Paper I).
  \vspace{0.1cm}
  \item[$\bullet$] Whenever detected, their progenitors were heavily obscured in the optical and NIR domain, but bright sources in the mid-infrared. These objects were associated with dust-enshrouded, extreme AGB stars in the range of 8-12 M$_{\odot}$ \citep{Thompson2009}.
  \vspace{0.1cm}
  \item[$\bullet$] When monitored for several years, ILRTs decline well below the progenitor level before disappearing, even in the MIR domain (\citealt{Adams2016}, Paper I). This favours the interpretation of ILRTs as terminal events.
  \vspace{0.1cm}
  \item[$\bullet$] Their spectra evolve slowly and are characterized by low excitation, narrow lines, with no signs of broad features even 17 months after the discovery.
\end{itemize}

\textbf{Eruption of an intermediate-mass star} -- A first scenario proposed for the formation of ILRTs was the non-terminal eruption of a $\sim$9-15 M$_{\odot}$ star, which reached its Eddington limit and became dynamically unstable after shedding a considerable amount of mass \citep{Hump2011}. The similarity between NGC 300 OT and the yellow hypergiant IRC+10420 (\citealt{Bond2009,Berger2009,Hump2011}) corroborated the idea of a stellar outburst producing an optically thick expanding envelope. However, this scenario struggles to explain the disappearance of the star observed years later in the mid infrared domain, since in the original view the progenitor would survive the eruption.

\textbf{Stellar mergers} -- Thanks to the careful analysis of V1309 Sco (e.g. \citealt{TylendaV1309Sco,MasonEV1309Scorpii}) the class of transients known as red novae was soundly associated with non-degenerate stellar mergers. 
Depending on the masses of the progenitor stars, the peak magnitude of red novae can span from being significantly fainter than a classical nova (M$_{r}$$\sim$--4.5 mag, e.g. OGLE-2002-BLG-360, \citealt{Pasto2021blu_OGLE}) up to competing with genuine core collapse SNe (M$_{r}$$\sim$--15.5 mag, e.g. AT 2017jfs, \citealt{TylendaOGLE}). On the other hand, ILRTs known to date peak consistently in the range --15 mag $\lesssim$ M$_{r}\lesssim$--11 mag (Paper I, \citealt{Cai2021ILRTs}).
Thus, there is some overlap in the energy radiated by ILRTs and the brightest red novae: this degeneracy is worsened by the similar spectral features shown by ILRTs and red novae during their photospheric phase. This can give rise to ambiguous objects, for which a clear classification is uncertain (e.g. M85 OT2006-1, \citealt{KulkarniM85,Pasto2007M85}). At the same time, there are some peculiarities in these two classes of transients that help with their identification: several (although not all) red novae present a double-peaked light curve, typically weaker or absent [Ca II] emission lines, and most importantly red novae always display molecular absorption bands in their late time spectra \citep{PastoGapTransients}, while ILRTs never show such bands. Furthermore, the progenitor systems of different red novae were observed in the optical domain (e.g. \citealt{TylendaV1309Sco,PastoRedNovae2019}), excluding the presence of large dust quantities along the line of sight, which is instead a key feature of ILRTs progenitors. In principle, these differences could be tentatively explained with a different viewing angle on the same type of progenitor system (such as binary stars surrounded by a dust torus): it was indeed suggested that the same physical mechanism could power both ILRTs and red novae (e.g. \citealt{Soker2021Jets}). However, the expected result of a stellar merger struggles to be reconciled with the severe mid infrared dimming observed for ILRTs over the course of a few years. Prolonged monitoring is therefore crucial to confirm the complete disappearance of the object, since there are known cases of transients settling on a lower luminosity compared to their progenitor (e.g. \citealt{JWST_stopDimming}).

\textbf{Failed supernovae} -- A hypothetical link was suggested between ILRTs and fallback supernovae, events in which a large mass fraction of the exploding star in unable to escape the gravitational potential of the compact object formed during the core collapse (e.g. \citealt{Takashi2010_Fallback}). \cite{Tsuna2020_ILRT_BH} found that the low energy radiated and the low photospheric temperature reached by ILRTs is compatible with radiation hydrodynamical simulations of matter ejected by a failed supernova which is interacting with pre-existing CSM. This interpretation has the merit of linking together a terminal event with the low luminosity and the narrow lines spectra of these transients. However, as the authors remarked, their simulations involve Wolf-Rayet and blue supergiants, which are not compatible with the progenitors of ILRTs discovered so far. Additionally, the monitoring of NGC 300 OT and SN 2008S spanning several years has not shown signs of fallback accretion on the black hole, which would be expected in this scenario.

\textbf{Electron capture supernovae} -- Multiple works have proposed the association between ILRTs and ECSNe \citep{Botticella2009,Pumo2009,CaiAT2017be,MaxStritzi2020ILRT,Cai2021ILRTs}. The 8-12 M$_{\odot}$ extreme AGB progenitors identified by \cite{Thompson2009} are the ideal candidates for giving rise to an ECSN explosion, and the intrinsic weakness of the event expected from the theory \citep{Kitaura2006} is encouraging when considering these underluminous objects. ILRTs, despite showing some degree of individuality, appear to be a relatively homogeneous family of objects, especially compared with the profound variability of the observable parameters within the red novae class. This can be interpreted as transients arising from similar progenitors (8-12 M$_{\odot}$ extreme AGB stars) whose differences are caused by the interaction with different environments.
The declining of the transient below the progenitor level is definitely expected in this scenario, given the terminal nature of ECSNe. In this context, the late time spectra characterized by narrow emission lines which were shown in this work deserve additional commentary. Type IIn SNe have shown that large quantities of CSM can bottle up the ejecta of violent stellar explosions for several years (at least 4 years in the case of SN 2015da; \citealt{Tartaglia2015da}). Actually, among SNe IIn it is rare to witness the emergence of the ejecta from the CSM, as observed for example for SN 1995N years after its explosion \citep{Fransson1995N, Pasto1995N}. In analogy, the absence of ejecta signatures in our latest ILRTs spectra does not exclude their presence underneath the CSM. The ECSN origin of these transients therefore remains a valid interpretation of the observed data, also corroborated by indirect indicators such as event rates (\citealt{Cai2021ILRTs}, \citealt{Karambelkar2023}) and dust chemical composition (\citealt{Rose_19abn_}).

\section{Summary and Conclusions}

We analysed a rich collection of optical and NIR spectra of four ILRTs. Their slow evolution coupled with the presence of narrow (few 10$^{2}$ km s$^{-1}$) emission lines is indicative of the presence of optically thick CSM surrounding the transients. Balmer lines as well as Ca features (Ca H\&K absorption doublet, [Ca II] doublet and Ca II NIR triplet) dominate the optical spectra and are recognizable at most epochs.
The O I $\lambda$ 8446 line appearing at late time is a consequence of the Bowen fluorescence, and could be the result of coupling between the O I $\lambda$1025.76 line in the CSM and the Ly$\beta$ line emitted by underlying ejecta.
The luminosity evolution of the H emission lines and Ca II NIR triplet display a characteristic double peaked behaviour, an opposite trend compared with the neighbouring continuum flux: this is possibly linked to the interaction between fast ejecta and the slow moving CSM. 

The high resolution spectrum of NGC 300 OT obtained with VLT+UVES shows the presence of a complex environment, possibly multiple shells of gas, causing narrow absorption features to appear on the most prominent emission lines. The characteristic shape of the [Ca II] lines is likely a result of electron scattering in an expanding medium, and allows us to postulate the presence of an extended, slow ($\sim$30 km s$^{-1}$) wind surrounding the progenitor star.
Late time spectra of NGC 300 OT and AT 2019ahd display broad features never observed in any ILRT so far. We find a striking similarity with the behaviour of the red nova V1309 Sco, and we tentatively explain these broad emission components with a blend of several narrow lines, possibly Fe transitions. 
In Paper I we discussed how the low energy released by ILRTs coupled with their terminal nature (at the very least for NGC 300 OT) make them appealing ECSN candidates. The spectroscopic analysis performed in this work provides a characterisation of the CSM surrounding these transients: while we do not find a smoking gun evidence for the explosion scenario, our findings are overall consistent with a weak ECSN event within a thick CSM.


\begin{acknowledgements}
We thank Steve Shore and Valentina D'Orazi for the useful and lively discussions.
We thank the staff of the various observatories where data were obtained for their assistance.
Based on observations made with the Nordic Optical Telescope, owned in collaboration by the University of Turku and Aarhus University, and operated jointly by Aarhus University, the University of Turku and the University of Oslo, representing Denmark, Finland and Norway, the University of Iceland and Stockholm University at the Observatorio del Roque de los Muchachos, La Palma, Spain, of the Instituto de Astrofisica de Canarias.
Observations from the Nordic Optical Telescope were obtained through the NUTS2 collaboration which are supported in part by the Instrument Centre for Danish Astrophysics (IDA) and the Finnish Centre for Astronomy with ESO (FINCA) via Academy of Finland grant nr 306531. The data presented here were obtained in part with ALFOSC, which is provided by the Instituto de Astrofisica de Andalucia (IAA).
Time-domain research by the University of Arizona team, and D.J.S. is supported by NSF grants AST-1821987, 1813466, 1908972, 2108032, and 2308181, and by the Heising-Simons Foundation under grant \#2020-1864.
H.K. was funded by the Research Council of Finland projects 324504, 328898, and 353019.
This work makes use of data from the Las Cumbres Observatory network. The LCO team is supported by NSF grants AST--1911225 and AST--1911151, and NASA SWIFT grant 80NSSC19K1639. 
Data were also obtained at the Liverpool Telescope, which is operated on the island of La Palma by Liverpool John Moores University in the Spanish Observatorio del Roque de los Muchachos with financial support from the UK Science and Technology Facilities Council. 
Based on observations made with the Gran Telescopio Canarias (GTC), installed in the Spanish Observatorio del Roque de los Muchachos of the Instituto de Astrofísica de Canarias, in the island of La Palma.
Based on observations collected at the European Organisation for Astronomical Research in the Southern Hemisphere, Chile, as part of ePESSTO+ (the advanced Public ESO Spectroscopic Survey for Transient Objects Survey).
ePESSTO+ observations were obtained under ESO program IDs 1103.D-0328 and 106.216C (PI: Inserra).
LCO data have been obtained via OPTICON. The OPTICON project has received funding from the European Union’s Horizon 2020 research and innovation programme under grant no. 730890.
Time domain research by the University of Arizona team and D.J.S. \ is supported by NSF grants AST-1821987, 1813466, 1908972, \& 2108032, and by the Heising-Simons Foundation under grant \#20201864. 
G.V., N.E.R. and I.S. acknowledge financial support from PRIN-INAF 2022 ``Shedding light on the nature of gap transients: from the observations to the models".
A.R. acknowledges financial support from the GRAWITA Large Program Grant (PI P. D’Avanzo) and the PRIN-INAF 2022 \textit{"Shedding light on the nature of gap transients: from the observations to the models}.
Y.-Z. Cai is supported by the National Natural Science Foundation of China (NSFC, Grant No. 12303054), the Yunnan Fundamental Research Projects (Grant No. 202401AU070063) and the International Centre of Supernovae, Yunnan Key Laboratory (No. 202302AN360001).
L.G. acknowledges financial support from the Spanish Ministerio de 
Ciencia e Innovaci\'on (MCIN), the Agencia Estatal de Investigaci\'on 
(AEI) 10.13039/501100011033, and the European Social Fund (ESF).
This work was funded by ANID, Millennium Science Initiative, ICN12\_009
M.N. is supported by the European Research Council (ERC) under the European Union’s Horizon 2020 research and innovation programme (grant agreement No.~948381). Part of the funding for GROND (both hardware as well as personnel) was generously granted from the Leibniz-Prize to Prof. G. Hasinger (DFG grant HA 1850/28-1).
H.K. was funded by the Academy of Finland projects 324504 and 328898. 
R.K. acknowledges support from the Research Council of Finland (340613).



\end{acknowledgements}

%
%
\bibliographystyle{aa}
\bibliography{example}

\section*{Affiliations}
\begin{small}
\textit{$^{1}$INAF -- Osservatorio Astronomico di Padova, Vicolo 
             dell’Osservatorio 5, I-35122 Padova, Italy\\
$^{2}$INAF -- OATS, Via G.B. Tiepolo 11, 34143 Trieste, Italy\\
$^{3}$INAF -- Osservatorio Astronomico di Brera, Via E. Bianchi 46, 23807, Merate (LC), Italy \\
$^{4}$Yunnan Observatories, Chinese Academy of Sciences, Kunming 650216, P.R. China \\
$^{5}$International Centre of Supernovae, Yunnan Key Laboratory, Kunming 650216, P.R. China \\
$^{6}$Key Laboratory for the Structure and Evolution of Celestial Objects, Chinese Academy of Sciences, Kunming 650216, P.R. China \\
$^{7}$Graduate Institute of Astronomy, National Central University, 300 Jhongda Road, 32001 Jhongli, Taiwan \\
$^{8}$SRON, Netherlands Institute for Space Research, Niels Bohrweg 4, 2333 CA, Leiden, The Netherlands \\
$^{9}$Department of Astrophysics/IMAPP, Radboud University Nĳmegen, P.O. Box 9010, 6500 GL, Nĳmegen, The Netherlands \\
$^{10}$Institute of Space Sciences (ICE, CSIC), Campus UAB, Carrer de Can Magrans s/n, E-08193 Barcelona, Spain \\
$^{11}$School of Physics, O’Brien Centre for Science North, University College Dublin, Belfield, Dublin 4, Ireland \\
$^{12}$The Oskar Klein Centre, Department of Astronomy, Stockholm University, AlbaNova, SE-10691 Stockholm, Sweden \\
$^{13}$Hiroshima Astrophysical Science Center, Hiroshima University, Japan \\
$^{14}$Department of Physics, Florida State University, 77 Chieftan Way, Tallahassee, FL 32306, USA \\
$^{15}$Las Cumbres Observatory, 6740 Cortona Dr. Suite 102, Goleta, CA, 93117, USA\\
$^{16}$Department of Physics, University of California, Santa Barbara, CA, 93106, USA\\
$^{17}$Cardiff Hub for Astrophysics Research and Technology, School of Physics \& Astronomy, Cardiff University, Queens Buildings, The Parade, Cardiff, CF24 3AA, UK \\
$^{18}$INAF, Osservatorio Astronomico di Capodimonte, Salita Moiariello 16, I-80131 Napoli, Italy \\
$^{19}$DARK, Niels Bohr Institute, University of Copenhagen, Jagtvej 128, 2200 Copenhagen, Denmark \\
$^{20}$Caltech/IPAC, Mailcode 100-22, Pasadena, CA 91125, USA \\
$^{21}$Tuorla Observatory, Department of Physics and Astronomy, University of Turku, 20014, Turku, Finland \\
$^{22}$Department of Astronomy, AlbaNova University Center, Stockholm University, SE-10691 Stockholm, Sweden \\
$^{23}$Astrophysics Research Institute, Liverpool John Moores University, IC2, Liverpool Science Park, 146 Brownlow Hill, Liverpool L3 5RF, UK \\
$^{24}$Max-Planck-Institut f\"{u}r Astrophysik, Karl-Schwarzschild Str. 1, D-85748 Garching, Germany \\
$^{25}$Aryabhatta Research Institute of Observational Sciences (ARIES), Manora Peak, Nainital 263001 India \\
$^{26}$Instituto de Alta Investigaci\'on, Universidad de Tarapac\'a, Casilla 7D, Arica, Chile\\
$^{27}$School of Physics, Trinity College Dublin, College Green, Dublin 2, Ireland \\
$^{28}$Steward Observatory, University of Arizona, 933 North Cherry Avenue, Tucson, AZ 85721-0065, USA\\
$^{29}$Department of Physics, University of Oxford, Keble Road, Oxford, OX1 3RH \\
$^{30}$Astrophysics Research Centre, School of Mathematics and Physics, Queens University Belfast, Belfast BT7 1NN, UK \\
$^{31}$Department of Physics and Astronomy, Aarhus University, Ny Munkegade 120, DK-8000 Aarhus C, Denmark \\
$^{32}$INAF – Osservatorio Astronomico d’Abruzzo, via M. Maggini snc, Teramo, I-64100, Italy \\
$^{33}$Department of Physics, University of California, Davis, CA 95616, USA \\
$^{34}$ European Southern Observatory, Alonso de C\'ordova 3107, Casilla 19, Santiago, Chile \\
$^{35}$Millennium Institute of Astrophysics, Nuncio Monsenor S\'otero Sanz 100, Providencia, 8320000 Santiago, Chile \\
$^{36}$INAF-Osservatorio Astrofisico di Catania, Via Santa Sofia 78, I-95123 Catania, Italy \\
$^{37}$Instituto de Astrof\'{i}sica, Universidad Andres Bello, Fernandez Concha 700, Las Condes, Santiago RM, Chile \\
$^{38}$ICRANet, Piazza della Repubblica 10, I-65122 Pescara, Italy \\
$^{39}$Institut für Theoretische Physik, Goethe Universität, Max-von-Laue-Str. 1, 60438 Frankfurt am Main, Germany \\
$^{40}$INFN-TIFPA, Trento Institute for Fundamental Physics and Applications, Via Sommarive 14, I-38123 Trento, Italy \\
$^{41}$Institut d’Estudis Espacials de Catalunya (IEEC), E-08034 Barcelona, Spain. \\
$^{42}$Astronomical Observatory, University of Warsaw, Al. Ujazdowskie 4, 00-478 Warszawa, Poland \\
$^{43}$Center for Astrophysics, Harvard \& Smithsonian, Cambridge, Massachusetts, MA 02138, US \\
$^{44}$The NSF AI Institute for Artificial Intelligence and Fundamental Interactions, USA \\
$^{45}$Finnish Centre for Astronomy with ESO (FINCA), University of Turku, Väisäläntie 20, 21500 Piikkiö, Finland \\
$^{46}$DTU Space, National Space Institute, Technical University of Denmark, Elektrovej 327, 2800 \\
$^{47}$Istituto Nazionale di Astrofisica - Osservatorio Astronomico di Roma, Via Frascati 33, I-00040, Monteporzio Catone, Italy \\
$^{48}$Kavli Institute for the Physics and Mathematics of the Universe (WPI), The University of Tokyo Institutes for Advanced Study, The University of Tokyo, Kashiwa, Chiba 277-8583, Japan \\
$^{49}$INFN. Sezione di Perugia, via A. Pascoli s/n, I-06125 Perugia, Italy \\
$^{50}$Dipartimento di Fisica e Astronomia ``G. Galilei'', Universit\`{a} degli studi di Padova Vicolo dell’Osservatorio 3, I-35122 Padova, Italy \\
$^{51}$IAASARS, National Observatory of Athens, Metaxa \& Vas. Pavlou St., 15236, Penteli, Athens, Greece \\
$^{52}$Department of Astronomy, University of Virginia, Charlottesville, VA 22904, USA \\
$^{53}$Max-Planck-Institut f{\"u}r Extraterrestrische Physik, Giessenbachstra{\ss}e 1, 85748, Garching, Germany \\
$^{54}$Department of Physics and Astronomy, University of North Carolina at Chapel Hill, Chapel Hill, NC 27599, USA \\
$^{55}$Cosmic Dawn Center (DAWN), Denmark \\
$^{56}$Niels Bohr Institute, University of Copenhagen, Jagtvej 128, 2200 København N, Denmark \\
$^{57}$Manipal Centre for Natural Sciences, Manipal Academy of Higher Education, Manipal - 576104, Karnataka, India \\
$^{58}$Department of Physics and Astronomy, Johns Hopkins University, Baltimore, MD 21218, USA \\
$^{59}$Space Telescope Science Institute: Baltimore, MD, US \\
$^{60}$Indian Institute Of Astrophysics, 100 Feet Rd, Santhosapuram, 2nd Block, Koramangala, Bengaluru, Karnataka 560034, India \\
}
\end{small}

\section*{Appendix: Additional tables and figures} \label{data_log}
\setcounter{table}{0}
\renewcommand{\thetable}{A\arabic{table}}

\setcounter{figure}{0}
\renewcommand{\thefigure}{A\arabic{figure}}

\begin{table*}

\begin{adjustbox}{tabular= cccccc, center}

&  & \textbf{AT 2019abn} &  & \\
\\ \hline
  Date & Telescope+Instrument & Grism & Range (\AA ) & Res (\AA ) & Exp. time (s) \\ 
 \hline
 \hline
 & & &  & \\
 2019-01-26 &  WHT+LIRIS &  HK &  14000-24000 &  13 &  3840 \\
 2019-01-27 &  WHT+LIRIS &  zJ &  9000-15000 &  8 &  2160 \\
 2019-01-27 &  NOT+ALFOSC &  gr4 &  3200-9600 &  16 &  1800 \\
 2019-01-29 &  WHT+ACAM &  V400 &  4000-9200 &  14 &  900 \\
 2019-01-30 &  WHT+LIRIS &  zJ &  9000-15000 &  8 &  1440 \\
 2019-01-30 &  WHT+ACAM &  V400 &  4000-9200 &  14 &  900 \\
 2019-02-01 &  WHT+ACAM &  V400 &  4000-9200 &  14 &  900 \\
 2019-02-06 &  NOT+ALFOSC &  gr3 &  4000-7000 &  12 &  2400 \\
 2019-02-11 &  NOT+ALFOSC &  gr4 &  3200-9600 &  16 &  2800 \\
 2019-02-20 &  NOT+ALFOSC &  gr4 &  3200-9600 &  12 &  3600 \\
 2019-02-23 &  NOT+ALFOSC &  gr4 &  4000-9600 &  16 &  2400 \\
 2019-02-28 &  NOT+ALFOSC &  gr4 &  3200-9600 &  12 &  2400 \\
 2019-03-02 &  LCO+FLOYDS &  red/blue &  3500-10000 &  13 &  3600 \\
 2019-03-15 &  NOT+ALFOSC &  gr4 &  3200-9600 &  12 &  2700 \\
 2019-03-18 &  LCO+FLOYD &  red/blue &  3500-10000 &  11 &  3600 \\
 2019-03-23 &  NOT+ALFOSC &  gr4 &  3200-9600 &  12 &  2700 \\
 2019-03-29 &  LCO+FLOYD &  red/blue &  3500-10000 &  13 &  3600 \\
 2019-04-04 &  CAHA+PMAS &  -- &  4500-7300 &  9 &  600 \\
 2019-04-09 &  LCO+FLOYD &  red/blue &  3500-10000 &  13 &  3600 \\
 2019-04-23 &  NOT+ALFOSC &  gr4 &  3200-9600 &  12 &  1800 \\
 2019-05-09 &  LCO+FLOYD &  red/blue &  3500-10000 &  13 &  3600 \\
 2019-05-15 &  NOT+ALFOSC &  gr4 &  3200-9600 &  15 &  3600 \\
 2019-05-20 &  LCO+FLOYD &  red/blue &  3500-10000 &  11 &  3600 \\
 2019-06-04 &  NOT+ALFOSC &  gr4 &  3200-9600 &  14 &  3600 \\
 2019-06-22 &  LCO+FLOYD &  red/blue &  3500-10000 &  13 &  3600 \\
 2019-07-20 &  NOT+ALFOSC &  gr4 &  3200-9600 &  12 &  3600 \\
 2019-07-26 &  LCO+FLOYD &  red/blue &  3500-10000 &  12 &  3600 \\
 2019-08-20 &  NOT+ALFOSC &  gr4 &  3200-9600 &  15 &  3600 \\
 2019-08-30 &  NOT+ALFOSC &  gr4 &  3200-9600 &  12 &  3600 \\

 &  &  &   & \\
 \hline

\end{adjustbox}
\caption{Log of spectroscopic observations for AT 2019abn.}
\label{SpecLog2019abn}
\end{table*}


\begin{table*}

\begin{adjustbox}{tabular= cccccc, center}
&  & \textbf{AT 2019ahd} &  & \\
\\ \hline
  Date & Telescope+Instrument & Grism & Range (\AA ) & Res (\AA ) & Exp. time (s) \\ 
 \hline
 \hline
 & & &  &\\
 2019-01-30 & SALT+RSS & PG0900 & 3500-9300 & 5 & 1890 \\
 2019-02-01 & WHT+ACAM & V400 & 4000-9200 & 11 & 1800 \\
 2019-02-06 & NOT+ALFOSC & gr4 & 3200-9600 & 12 & 3600 \\
 2019-02-11 & NOT+ALFOSC & gr4 & 3200-9600 & 16 & 3600 \\
 2019-02-28 & NOT+ALFOSC & gr4 & 3200-9600 & 12 & 3600 \\
 2019-03-05 & SALT+RSS & PG0700 & 3500-9300 & 16 & 1800 \\
 2019-03-15 & NOT+ALFOSC & gr4 & 3200-9600 & 12 & 2700 \\
 2019-04-07 & NOT+ALFOSC & gr4 & 3200-9600 & 12 & 3000 \\
 2019-04-18 & NOT+ALFOSC & gr4 & 3200-9600 & 12 & 3600 \\
 2019-05-14 & NOT+ALFOSC & gr4 & 3200-9600 & 12 & 3600 \\
 2019-05-21 & SALT+RSS & PG0700 & 3500-9300 & 15 & 1800 \\
 2019-11-23 & GTC+OSIRIS & R1000R & 5100-10300 & 6 & 1350 \\

 & & &  &\\
 \hline

\end{adjustbox}

\caption{Log of spectroscopic observations for AT 2019ahd.}
\label{SpecLog2019ahd}
\end{table*}


\begin{table*}
\caption{Log of spectroscopic observations for AT 2019udc.}
\label{SpecLog2019udc}
\begin{adjustbox}{tabular= cccccc, center}

&  & \textbf{AT 2019udc} &  & \\
\\ \hline
  Date & Telescope+Instrument & Grism & Range (\AA ) & Res (\AA ) & Exp. time (s) \\ 
 \hline
 \hline
 & & &  &\\
 2019-11-04 & LCO+FLOYDS & red/blue & 3500-10000 & 12 & 2700 \\
 2019-11-04 & HCT-HFOSC2 & gr7 & 4000-7600 & 18 & 1800 \\
 2019-11-05 & HCT-HFOSC2 & gr7 & 4000-7600 & 18 & 2700 \\
 2019-11-06 & LCO+FLOYDS & red/blue & 3500-10000 & 18 & 2700 \\
 2019-11-13 & LCO+FLOYDS & red/blue & 3500-10000 & 12 & 3600 \\
 2019-11-14 & LCO+FLOYDS & red/blue & 3500-10000 & 12 & 3600 \\
 2019-11-15 & LCO+FLOYDS & red/blue & 3500-10000 & 16 & 3600 \\
 2019-11-17 & LCO+FLOYDS & red/blue & 3500-10000 & 12 & 3600 \\
 2019-11-18 & LCO+FLOYDS & red/blue & 3500-10000 & 15 & 3600 \\
 2019-11-22 & GTC+OSIRIS & R1000R+B & 3600-10300 & 7 & 1380 \\
 2019-11-30 & LCO+FLOYDS & red/blue & 3500-10000 & 12 & 3600 \\
 2019-12-03 & LCO+FLOYDS & red/blue & 3500-10000 & 12 & 3600 \\
 2019-12-06 & Baade+FIRE & prism & 8000-22000 & 30 & 888 \\
 2019-12-14 & NOT+ALFOSC & gr4 & 3200-9600 & 20 & 2800 \\
 2019-12-24 & LCO+FLOYDS & red/blue & 3500-10000 & 13 & 3600 \\
 2019-12-26 & NTT+EFOSC2 & gr20 & 6000-7100 & 13 & 2400 \\
 2020-01-19 & LCO+FLOYDS & red/blue & 3500-10000 & 12 & 3600 \\
 & & &  &\\
 \hline

\end{adjustbox}

\end{table*}



\begin{table*}
\caption{Log of spectroscopic observations along with main features of the instruments used. FORS spectra were taken from \protect \cite{PatatJulySpec}.}
\label{SpecLog}
\begin{adjustbox}{tabular= cccccc, center}

&  & \textbf{NGC 300 OT} &  & \\
\\ \hline
  Date & Telescope+Instrument & Grism & Range (\AA ) & Res (\AA ) & Exp. time (s) \\ 
 \hline
 \hline
 & & &  &\\
 2008-05-16 & BMagellan+IMACSshort & g200 & 4000-10500 & 6 & 100x2\\
 2008-05-30 & DuPont+B\&C & g300 & 3500-9600 & 8 & 300\\
 2008-06-04 & DuPont+B\&C & g300 & 3500-9600 & 8 & 300\\
 2008-06-06 & NTT+EFOSC & gm3-gm5 & 3300-9200 & B11 R15 & 1460x2\\
 2008-06-11 & CMagellan+LDSS3 & - & 3700-9800 & 4 & 300x3\\
 2008-06-13 & NTT+EFOSC & gm3-gm5 & 3300-9200 & B11 R15 & 1460x2\\
 2008-06-15 & NTT+EFOSC & gm3-gm5 & 3300-9200 & B11 R15 & 1460x2\\
 2008-06-16 & BMagellan+IMACSlong & g300 & 3700-7900 & 4 & 300x3\\
 2008-07-01 & VLT+FORS & 300V & 3300-9200 & 11 & 1800 \\
 2008-07-02 & VLT+UVES & SHP700+HER\_5 & 3400-8000 & B0.12 R0.36 & 1500x6 \\
 2008-07-05 & NTT+EFOSC & gm3-gm5 & 3300-9200 & B11 R15 & 1460x2\\
 2008-07-08 & VLT+FORS & 300V & 3300-9200 & 11 & 2160 \\
 2008-07-15 & NTT+EFOSC & gm3-gm5 & 3300-9200 & B11 R15 & 1460x2\\
 2008-07-23 & CTIO4m+R-CSpec & KPGL3-1 & 4000-7600 & 5 & 720\\
 2008-07-24 & CTIO4m+R-CSpec & KPGL3-1 & 4000-7600 & 5 & 1200\\
 2008-07-25 & CTIO4m+R-CSpec & KPGL3-1 & 4000-7600 & 5 & 1200\\
 2008-08-05 & NTT+EFOSC & gm3-gm5 & 3300-9200 & B11 R15 & 1460x2\\
 2008-09-17 & CMagellan+LDSS3 & VPH-All & 3700-9400 & 6 & 900\\
 2008-10-02 & Palomar200i+DBSP & 300/3990+158/7500 & 3300-10300 & B11 R16 & 1800x2\\
 2008-10-08 & TNG+NICS & IJHK & 8700-24700 & IJH16 K30 & -- \\
 2008-10-14 & WHT+ISIS & R158R+R300B & 3400-10200 & B4 R6 & 600x3 \\
 2009-01-07 & Gemini South+GMOS & - & 4000-8000 & 6 & 1580\\
 2009-01-21 & Gemini South+GMOS & - & 4000-8000 & 6 & 1800\\
 2009-10-12 & VLT+X-Shooter & - & 5600-24400 & B0.7 R2.7 & 300 \\
 & & &  &\\
 \hline

\end{adjustbox}
\end{table*}

\begin{figure*}[ht]
\begin{center}
  \makebox[1.6\columnwidth][c]{\includegraphics[width=0.95\textwidth]{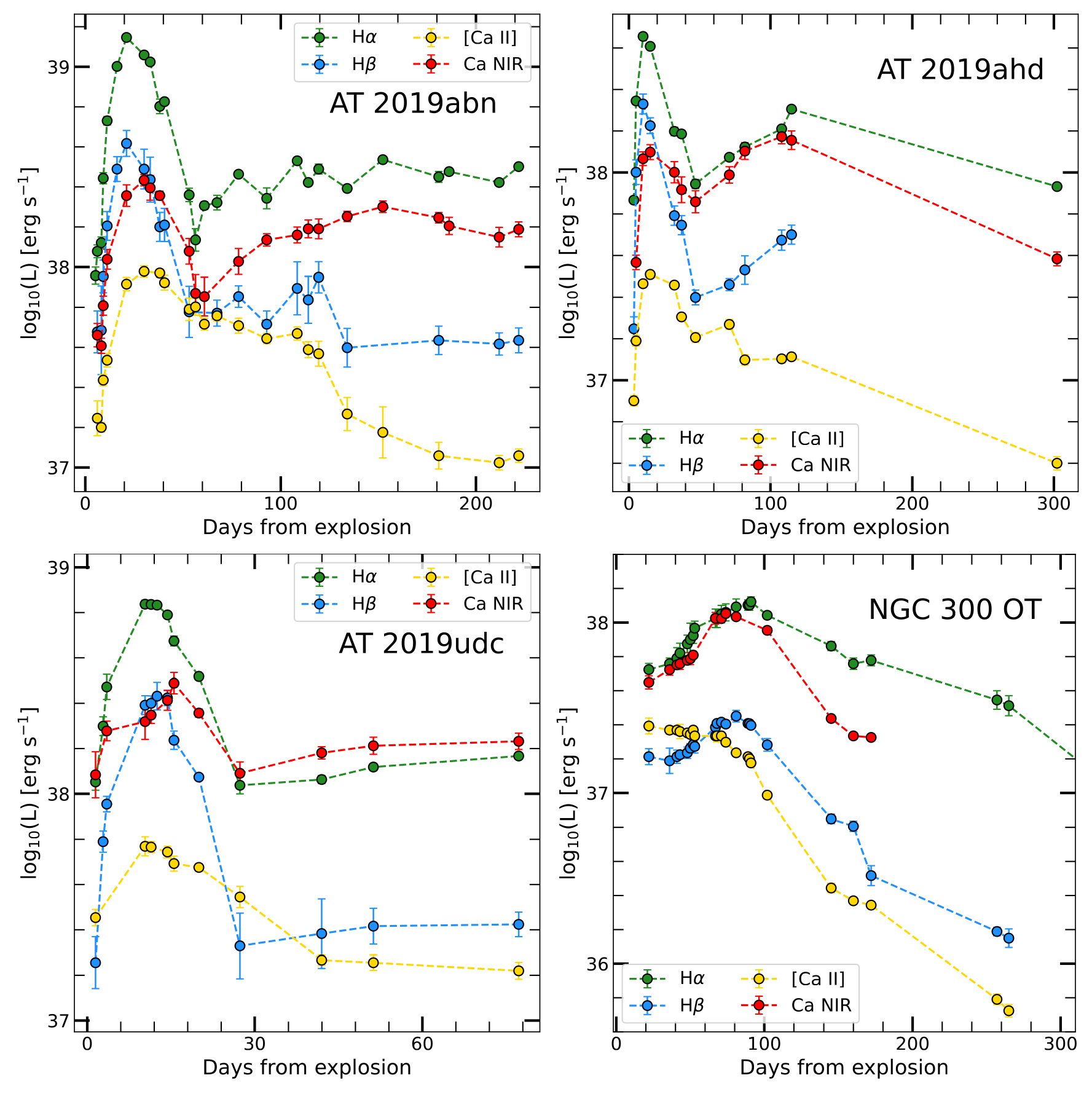}}%
  \caption{Luminosity of the emission lines for our sample of ILRTs, grouped by transient rather than by emission feature (as in Figure \ref{LineLuminosity}).}
  \label{LinesSameObject}
  \end{center}
  
\end{figure*}

\begin{figure*}[htb!]
\begin{center}
  \makebox[1.6\columnwidth][c]{\includegraphics[width=0.95\textwidth]{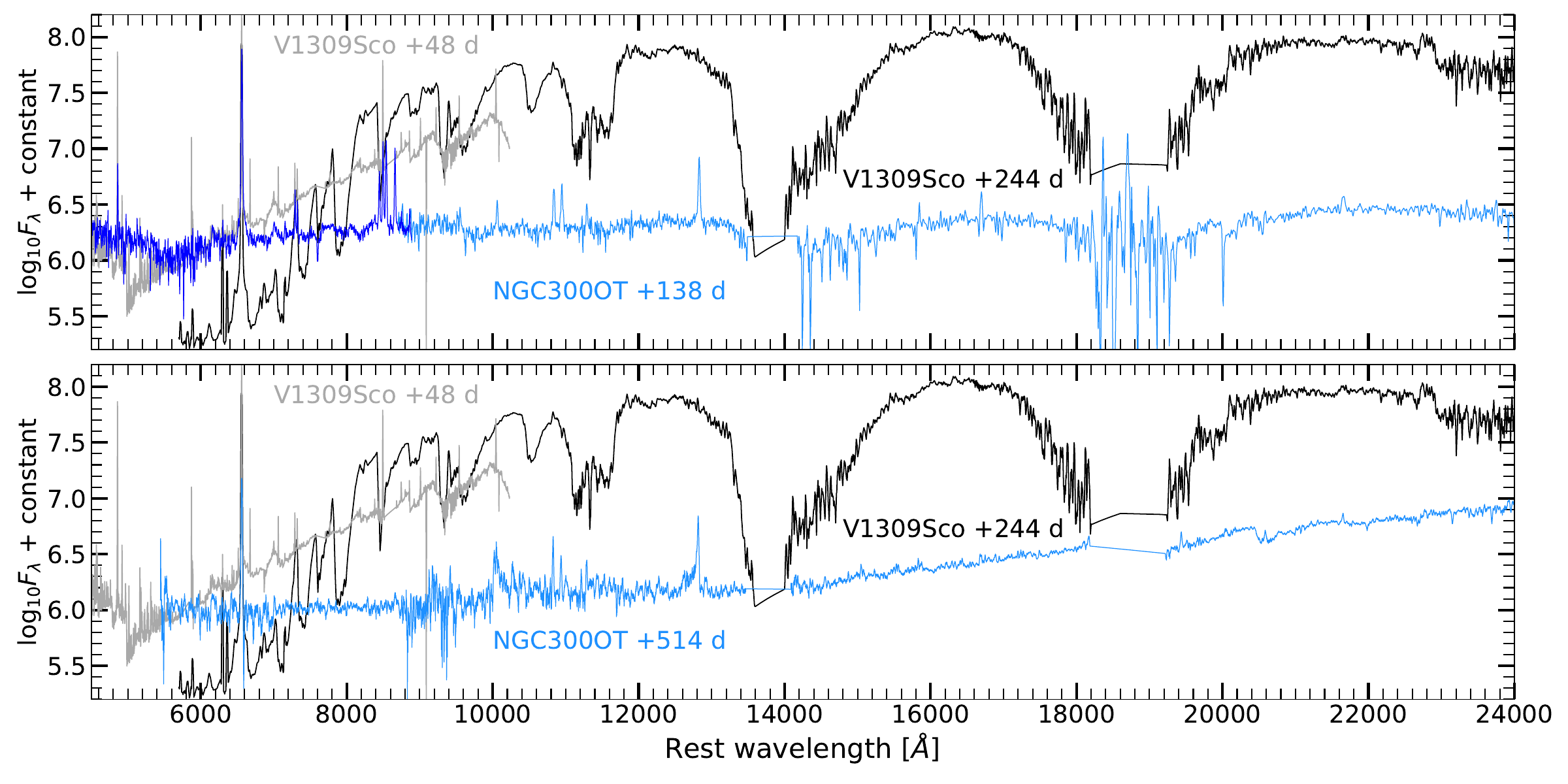}}%
  \caption{Late time NIR spectra of NGC 300 OT compared with V 1309 Sco. In the upper panel, the NIR spectrum of NGC 300 OT taken with TNG+NICS on 2008-10-08 (shown in light blue) is plotted along with the optical spectrum obtained with WHT+ISIS six days later (in blue).}
  \label{MolecularBandsNIR}
  \end{center}
  
\end{figure*}

\end{document}